\documentstyle[prc,aps,epsfig]{revtex}

\def\x{\mbox{\boldmath $x$}}
\def\z{\mbox{\boldmath $z$}}
\def\s{\mbox{\boldmath $s$}}
\def\h{\mbox{\boldmath $h$}}
\def\S{\mbox{\boldmath $S$}}
\def\N{\mbox{\boldmath $N$}}
\def\L{\mbox{\boldmath $L$}}

\def\p{\mbox{\boldmath $p$}}
\def\n{\mbox{\boldmath $n$}}
\def\P{\mbox{\boldmath $P$}}
\def\q{\mbox{\boldmath $q$}}

\def\N{\mbox{\boldmath $N$}}

\begin{document}
\draft
\title{Polarization observables in (${\vec \gamma},{\vec N}N$) 
reactions}
\author{C.~Giusti and F.~D.~Pacati}
\address{Dipartimento di Fisica Nucleare e Teorica dell'Universit\`a, 
Pavia\\
and Istituto Nazionale di Fisica Nucleare, Sezione di Pavia, Italy}
\maketitle
\begin{abstract}
The formalism of ($\vec{\gamma},\vec{N}N$) 
reactions is given where the 
incident photon is polarized and the outgoing nucleon polarization is detected. 
Sixteen structure functions and fifteen polarization observables are found in 
the general case, while only eight structure functions and seven polarization
observables survive in coplanar kinematics. Numerical examples are presented for 
the $^{16}$O($\gamma,np$) and $^{16}$O($\gamma,pp$) reactions. The transitions 
to the ground state of $^{14}$C and $^{14}$N are calculated in a model where 
realistic short-range and tensor correlations are taken into account for the 
$np$ pair, while short-range and long-range correlations are included in a 
consistent way for $pp$ pairs. The effects of the one-body and two-body 
components of the nuclear current and the role of correlations in cross sections 
and polarizations are studied and discussed.
\end{abstract}
%
\section{Introduction}
\label{sec:1}
For a long time electromagnetically induced two-nucleon knockout has been 
considered the most direct tool for exploring the properties of nucleon pairs
within nuclei and their interaction at short distance \cite{Gottfried,Oxford}. 
The cross section of an exclusive reaction contains the two-hole spectral 
function, which gives information on nuclear structure and correlations. 

Two nucleons can be naturally ejected by two-body currents, which effectively 
take into account the influence of subnuclear degrees of freedom like mesons 
and isobars. Direct insight into dynamical short-range correlations (SRC) can 
be obtained from the process where the real or virtual photon hits, through a 
one-body current, either nucleon of a correlated pair and both nucleons are 
then ejected from the nucleus. A reliable and consistent treatment of these 
two competing processes, both produced by the exchange of mesons between 
nucleons, is needed. Their role and relevance, however, can be different in 
different reactions and kinematics. It is thus possible to envisage situations 
where either process is dominant and various specific effects can be 
disentangled and separately investigated.

First pioneering ($\gamma,NN$) experiments were performed in Bonn \cite{Bonn} 
and Tokio \cite{Tokio} with poor statistical accuracy and energy resolution. 
Further experiments carried out at MAMI achieved much better 
results \cite{Mainz}. These experiments, with an energy resolution of about 6-9 
MeV, were able to resolve the major shells from which the two nucleons are 
emitted. The first high-resolution $^{16}$O($\gamma,np$)$^{14}$N experiment 
able to separate final states was performed at MAXlab in Lund \cite{Lund}. 
A high-resolution experiment for the $^{16}$O($\gamma,np$)$^{14}$ and 
$^{16}$O($\gamma,pp$)$^{14}$C reactions in the photon-energy range between 90 
and 270 MeV, aiming not only at separating final states but also at determining 
the momentum distributions of each state, has been recently approved in 
Mainz \cite{Mainz1}. Results for a first test and feasibility study are
presented in ref. \cite{Grab}.

Reaction calculations performed with different microscopic models based on a
two-nucleon knockout framework \cite{gnn,Gent2} confirm that ($\gamma,np$) and
($\gamma,pp$) reactions are generally dominated by two-body currents. Thus, 
they are useful to study the nuclear current, whose various components can be 
emphasized or suppressed in different conditions and kinematics. However, they 
are also sensitive to the other theoretical ingredients and represent a 
suitable tool to study the conditions of nucleon pairs within a nucleus. 
In fact, the angular distributions of the $^{16}$O($\gamma,np$)$^{14}$N and 
$^{16}$O($\gamma,pp$)$^{14}$C cross sections for transitions to different 
final states have a different shape, essentially determined by the components 
of the center-of-mass (c.m.) orbital angular momentum of the initial 
pair \cite{gnn}, which can be explored by comparing the theoretical predictions
with the experimental data. 

Interesting and complementary information is available from ($e,e'NN$) and 
($\gamma,NN$) reactions. 
First measurements of the exclusive $^{16}$O($e,e'pp$)$^{14}$C reaction have 
been performed at NIKHEF in Amsterdam \cite{Gerco,NIKHEF,Ronald} and MAMI in 
Mainz \cite{Rosner}. Investigations on these data 
\cite{Gerco,NIKHEF,Ronald,Rosner,sf} indicate that resolution of discrete 
final states provides an interesting tool to disentangle the two reaction
mechanisms due to one-body currents, and thus to SRC, and to two-body currents. 

Recent calculations \cite{pn} have shown that also the cross sections of the 
exclusive $^{16}$O($e,e'np$)$^{14}$N reaction are sensitive to details of the 
nuclear correlations and in particular to the presence of the tensor component. 

It is clear, however, that 
($e,e'np$) experiments are more difficult, as they require a triple coincidence 
measurement with the detection of a neutron. Thus, no data are available at 
present, but a proposal for the experimental study of the exclusive 
$^{16}$O($e,e'np$)$^{14}$N reaction has been approved in Mainz \cite{MAMI}.

Good opportunities for a more complete investigation of the properties of 
different pairs of nucleons in nuclei and of their interaction at short 
distance are offered by polarization measurements. Reactions with 
polarized particles give access to a larger number of observables, hidden in the 
unpolarized case and whose determination can impose more severe constraints to 
theoretical models. Some of these observables are expected to be sensitive to 
the small components of the transition amplitudes, which are generally masked by 
the dominant ones in the unpolarized cross section and that often contain 
interesting information on subtle effects. 
In photoreactions, and in particular in ($\gamma,np$) reactions, the two-body 
contributions are always dominant in the cross section with respect to one-body 
ones. In order to study correlations, the polarization measurements are 
therefore necessary.

The asymmetry of the cross section for linearly polarized photons in 
($\gamma,pp$) and ($\gamma,np$) reactions was studied in 
refs. \cite{Gent2,BGPR}. Numerical results for exclusive $pp$ and $np$ knockout from $^{16}$O can be found 
in ref. \cite{gnn}. First measurements with polarized photon beams have been 
performed at LEGS on $^3$He \cite{LEGSHe} and $^{16}$O \cite{LEGSO} and at MAMI 
on $^{12}$C \cite{MacGre}. In these first experiments, however, the energy 
resolution was not enough to separate specific discrete final states of the
residual nucleus.

The nucleon recoil polarization in photon-induced and electron-induced 
two-nucleon knockout has been considered in refs. \cite{Gent2} and \cite{Gent1}, 
respectively. The general formalism of the 
($\vec{e},e'\vec{N}N$) reaction and some theoretical 
predictions for the exclusive 
$^{16}$O($\vec{e},e'\vec{p}p$)$^{14}$C knockout reaction 
are given in 
ref. \cite{pppol}. The results of these investigations indicate that a combined 
measurement of cross sections and polarization components would provide an 
interesting tool to disentangle the different reaction mechanisms and clarify 
the behaviour of nucleon pairs in the nuclear medium. No data are available at 
present, but a measurement of the nucleon recoil polarization seems reasonably 
within reach of available experimental facilities.

The general case of ($\vec{\gamma},\vec{N}N$) reactions, 
where the incident photon is linearly or circularly polarized and/or the 
outgoing nucleon polarization is detected is considered in this paper. Besides 
the cross section, 15 new observables are obtained in the most general 
kinematics. The formalism and the polarization observables are given in  
sect.~\ref{sec:2}. Some nu\-mer\-i\-cal re\-sults for the ex\-clu\-sive 
$^{16}$O($\vec{\gamma},\vec{n}p$)$^{14}$N$_{\mathrm{g.s.}}$ and 
$^{16}$O($\vec{\gamma},\vec{p}p$)$^{14}$C$_{\mathrm{g.s.}}$ reactions 
are presented and discussed in sects.~\ref{sec:3} and \ref{sec:4}, 
respectively. Conclusions are drawn in sect.~\ref{sec:5}.

\section{Polarization observables in 
($\vec{\gamma},\vec{\N}\N$) reactions}
\label{sec:2}

The cross section of a reaction, where an incident photon, with momentum $\q$ 
and energy $E_\gamma$, is absorbed by the nucleus and 
two nucleons are emitted, with 
momenta $\p'_1$ and $\p'_2$ and energies $E'_1$ and $E'_2$, can be written 
in terms of four structure functions $f_{\lambda\lambda'}$ as
\begin{eqnarray} 
\frac{{\rm d}^{3}\sigma}{{\rm d}\Omega_{1}{\rm d}\Omega_{2}
{\rm d}E'_{2}} = \frac {2 \pi^2 \alpha} {E_\gamma} \, K [ & f_{11} &
+ \Pi_{\mathrm c} f'_{11} - \Pi_{\mathrm t} (f_{1-1} 
\cos 2\phi_\gamma \nonumber \\
& - & \bar{f}_{1-1} \sin 2\phi_\gamma) ], \label{eq:cs}
\end{eqnarray}
where $\Pi_{\mathrm c}$ and $\Pi_{\mathrm t}$ are the degrees of circular and 
linear polarization of the photon, $\phi_\gamma$ is the angle of the photon 
polarization vector relative to the scattering plane, \textit{i.e.} the 
($\q, \p'_1$) plane, and 
\begin{equation}
K = |\p'_1|\, E'_1 \,|\p'_2| \,E'_2 \,f_{\mathrm{rec}} ,
\end{equation} 
with
\begin{equation}
f_{\mathrm{rec}}^{-1} = 1- \frac {E'_2 } {E_{\mathrm r}} \frac {\p'_2 \cdot 
\p_{\mathrm r}} {|\p'_2|^2},
\end{equation}
where $\p_{\mathrm r}$ and $E_{\mathrm r}$ are the momentum and energy of the 
residual nucleus. In our reference frame the $\z$ axis is taken parallel to 
$\q$ and the momentum $\p'_1$ lies in the ($\x, \z$) plane.

The structure functions $f_{\lambda\lambda'}$ are obtained from the components of
the hadron tensor $W^{ij}$, with $i,j = x,y$, and are given by all the 
symmetrical and
antisymmetrical independent combinations of these components, \textit{i.e.}
\begin{eqnarray}
&f_{11} = W^{xx} + W^{yy}, \quad & f'_{11} = -i \, (W^{xy} - W^{yx}), 
\nonumber \\
&f_{1-1} = W^{yy} - W^{xx}, \quad & \bar{f}_{1-1} = W^{xy} + W^{yx}.
\label{eq:fll}
\end{eqnarray}

If the polarization of the nucleon 1 outgoing in the scattering plane is 
detected, the structure functions can in general be written as 
\begin{eqnarray}
f_{\lambda\lambda'} &=& h^u_{\lambda\lambda'} + \hat{\s} \cdot 
\h_{\lambda\lambda'}, \nonumber \\
f'_{\lambda\lambda'} &=& h'^u_{\lambda\lambda'} + \hat{\s} \cdot 
\h'_{\lambda\lambda'}
\label{eq:hll}
\end{eqnarray}
where $\hat{\s}$ is the unit vector in the spin direction of the recoil nucleon.
Thus, when the polarization of the outgoing nucleon is considered, 16 structure 
functions appear: 4 spin independent, $h^{\mathrm{u}}_{\lambda\lambda'}$ and 
$h'^{\mathrm{u}}_{\lambda\lambda'}$, and 12 
spin dependent, $h^k_{\lambda\lambda'}$ and $h'^k_{\lambda\lambda'}$, 4 for 
each one of the 3 components of $\h_{\lambda\lambda'}$ and 
$\h'_{\lambda\lambda'}$.

The explicit expressions of these structure functions in terms of the 
components of the hadron tensor $W^{\mu\nu}_{\alpha\alpha'}$, where 
$\alpha$ and $\alpha'$ are the eigenvalues of the spin of the nucleon whose
polarization is considered, can be easily obtained from eq.~(\ref{eq:fll}), by 
simply substituting the quantities $W^{\mu\nu}$, wherever they appear, with the 
following expressions \cite{pppol}:
\begin{eqnarray}
W^{\mu\nu}_{++} + W^{\mu\nu}_{--} \,\,\,\, &  {\mathrm {for}} &  \,\,
{ k = u,} \nonumber \\
W^{\mu\nu}_{+-} + W^{\mu\nu}_{-+}   \,\,\,\, & {\mathrm {for}} & \,\, 
{ k = x,} \nonumber \\
i (W^{\mu\nu}_{+-} - W^{\mu\nu}_{-+}) \,\,\,\, & {\mathrm {for}} & 
\,\, { k = y,} \nonumber \\
W^{\mu\nu}_{++} - W^{\mu\nu}_{--}  \,\,\,\, & {\mathrm {for}} & \,\, 
{ k = z.}
\label{eq:wmn}
\end{eqnarray} 
Note that in ref. \cite{pppol} there is a misprint in the case $k~=~y$.

Usually, the quantities $\h_{\lambda\lambda'}$ and $\h'_{\lambda\lambda'}$ 
are projected onto the basis of unit vectors given by $\hat{\L}$\/ (parallel 
to ${\p}\/'_1$), $\hat{\N}$\/ (in the direction of ${\q}\/\times{\p}\/'_1$) 
and $\hat{\S} = \hat{\N}\times\hat{\L}$\/, and the structure functions are 
thus given for the components $k = N, L, S$.

When the outgoing nucleon polarization is not detected, the cross
section is summed over the spin quantum numbers of the outgoing nucleon and the
spin independent structure functions $h^{\mathrm{u}}_{\lambda\lambda'}$ go over
to the structure functions $f_{\lambda\lambda'}$ of the unpolarized
case \cite{Oxford,pppol}. Thus, in this case only four structure functions are
obtained, $f_{\lambda\lambda'}= 2h^{\mathrm{u}}_{\lambda\lambda'}$, and four 
independent quantities can be measured: the unpolarized cross section
\begin{equation}
\sigma_0 = \frac {2 \pi^2 \alpha} {E_\gamma}\, K \, 2 h^u_{11} ,
\label{eq:s0}
\end{equation} 
the circular asymmetry
\begin{equation}
A_{\mathrm c} = \frac {\sigma(+) - \sigma(-)} {\sigma(+) + \sigma(-)}
= \frac {h'^u_{11}} {h_{11}^u}\, ,
\label{eq:ac}
\end{equation}
where $\sigma(+)$ and $\sigma(-)$ are the cross sections for a circularly
polarized photon and $\Pi_{\mathrm t}$ = 0,
and the two independent linear asymmetries,
\begin{equation}
\Sigma_{\frac {\pi} {2}} = \frac {\sigma(0) - \sigma(\frac {\pi} {2})} 
{\sigma(0) + \sigma(\frac {\pi} {2})} = - \frac {h^u_{1-1}} 
{h_{11}^u}
\label{eq:sp2}
\end{equation}
and
\begin{equation}
\Sigma_{\frac {\pi} {4}} = \frac {\sigma(\frac {\pi} {4}) - 
\sigma(-\frac {\pi} {4})} 
{\sigma(\frac {\pi} {4}) + \sigma(-\frac {\pi} {4})} = 
 \frac {\bar {h}^u_{1-1}} 
{h_{11}^u}\, ,
\label{eq:sp4}
\end{equation}
where $\sigma(0)$, $\sigma(\frac {\pi} {2})$ and $\sigma(\frac {\pi} {4})$ are 
the cross sections for a linearly polarized photon in the directions 
$\phi_\gamma = 0, {\frac{\pi} {2}}$, and ${\frac {\pi} {4}}$, 
respectively, and $\Pi_{\mathrm c}$ = 0.

When the polarization of the outgoing nucleon is detected, the 12 new structure
functions $h^k_{\lambda\lambda'}$ and 12 corresponding polarization observables, 
4 for each direction $N$, $L$, and $S$, are produced. For an unpolarized photon, 
we have the components of the polarization 
\begin{equation}
P^k = \frac {\sigma(+1) - \sigma(-1)} {\sigma(+1) + \sigma(-1)}
= \frac {h^{k}_{11}} {h_{11}^u}\, ,
\label{eq:pk}
\end{equation}
where $\sigma(+1)$ ($\sigma(-1)$) are the cross sections with a nucleon
polarized parallel (antiparallel) to the component $\hat {s}^k$ of the spin
direction given in eq.~(\ref{eq:hll}).

If the photon is circularly polarized and $\Pi_{\mathrm t}$ = 0, we have the 
components of the polarization transfer coefficient 
\begin{eqnarray}
P^k_{\mathrm c} & = &\frac {\sigma(+,+1) - \sigma(+,-1) - \sigma(-,+1) + 
\sigma(-,-1)} {\sigma(+,+1) + \sigma(+,-1) + \sigma(-,+1) 
+ \sigma(-,-1)}\nonumber \\
& = & \frac {h'^{k}_{11}} {h_{11}^u}\, ,
\label{eq:pck}
\end{eqnarray}
where, here and in the following,  the first argument in $\sigma$ refers 
to the photon polarization 
and the second to the nucleon polarization in the direction $\hat {s}^k$.

For a linearly polarized photon and $\Pi_{\mathrm c}$ = 0, we have the components
of the two polarization asymmetries
\begin{eqnarray}
\Sigma^k_{\frac {\pi} {2}} & = &\frac {\sigma(0,+1) - \sigma(0,-1) - 
\sigma(\frac {\pi} {2},+1) + 
\sigma(\frac {\pi} {2},-1)} {\sigma(0,+1) + \sigma(0,-1) + 
\sigma(\frac {\pi} {2},+1) + \sigma(\frac {\pi} {2},-1)} \nonumber \\
& = & -\frac {h^{k}_{1-1}} {h_{11}^u}
\label{eq:sp2k}
\end{eqnarray}
and
\begin{eqnarray}
\Sigma^k_{\frac {\pi} {4}} & = & \frac {\sigma(\frac {\pi} {4},+1) - 
\sigma(\frac {\pi} {4},-1) - \sigma(-\frac {\pi} {4},+1) + 
\sigma(-\frac {\pi} {4},-1)} {\sigma(\frac {\pi} {4},+1) + 
\sigma(\frac {\pi} {4},-1) + 
\sigma(-\frac {\pi} {4},+1) + \sigma(-\frac {\pi} {4},-1)} \nonumber \\
& = & \frac {\bar {h}^{k}_{1-1}} {h_{11}^u}\, ,
\label{eq:sp4k}
\end{eqnarray}
where the first argument in $\sigma$ refers to the value of the angle
$\phi_\gamma$.

In the case of a coplanar kinematics, where both the emitted nucleons and the
photon lie in the same plane, owing to parity conservation combined with the
general properties of the hadron tensor, several structure functions vanish and 
only 8 of them survive: $h^u_{11}$, $h^{\mathrm N}_{11}$, 
$h^u_{1-1}$, $h^{\mathrm N}_{1-1}$, $h'^{\mathrm L,S}_{11}$ and 
$\bar{h}^{\mathrm L,S}_{1-1}$ and thus the quantities $\sigma_0$, 
$P^{\mathrm N}$, $\Sigma_{\frac {\pi} {2}}$, 
$\Sigma^{\mathrm N}_{\frac {\pi} {2}}$, $P^{\mathrm{L,S}}_{\mathrm c}$, and 
$\Sigma^{\mathrm {L,S}}_{\frac {\pi} {4}}$ can be measured. This result can be 
derived along the same lines as the one obtained in ref. \cite{pppol} in the 
more general case of the ($\vec{e},e'\vec{N}N$) reaction, and is similar to 
the result obtained for the ($\vec{\gamma},\vec{N}$) 
reaction \cite{Oxford}. 

When the nucleon spin is not detected, in coplanar kinematics the only 
dynamical variable with a nonvanishing component in the $y$ direction is the 
pseudovector $\q \times \p'_1$. As no pseudovectors are available with a 
component along $x$, the matrix elements of the hadron tensor $W^{xy}$ and 
$W^{yx}$ must vanish.
Therefore, $h'^u_{11}$ = $\bar{h}^u_{1-1}$ = 0 and,
consequently, $A_{\mathrm c}$ = $\Sigma_{\frac {\pi} {4}}$ = 0, and  only 
$\Sigma_{\frac {\pi} {2}}$ survives. 

When the nucleon spin is detected, owing to the properties of the  
spin-${\frac{1}{2}}$ matrix, only a linear dependence of the structure functions
on the spin is allowed [see eq.~(\ref{eq:hll})]. As the spin is a pseudovector 
and in order to fulfill parity conservation, it must be coupled in the hadron 
tensor with the only available pseudovector independent of the spin, 
\textit{i.e.} with 
$\q \times \p'_1$, which in coplanar kinematics is directed along $y$. 
Therefore, we have: $W^{xx}$ = 0, $W^{yy}$ dependent on $s^{\mathrm N}$, and 
$W^{xy}$ independent of $s^{\mathrm N}$. As a consequence, one has: 
$h'^{\mathrm N}_{11}$ = $\bar{h}^{\mathrm N}_{1-1}$ = $h^{\mathrm L,S}_{11}$ = 
$h^{\mathrm L,S}_{1-1}$ = 0. Thus, $P^{\mathrm L,S}$,
$\Sigma^{\mathrm L,S}_{\frac {\pi} {2}}$, $P^{\mathrm N}_{\mathrm c}$ and 
$\Sigma^{\mathrm N}_{\frac {\pi} {4}}$ vanish and only  $P^{\mathrm N}$,
$\Sigma^{\mathrm N}_{\frac {\pi} {2}}$, $P^{\mathrm L,S}_{\mathrm c}$ and 
$\Sigma^{\mathrm L,S}_{\frac {\pi} {4}}$ survive.

The behaviour of the hadron tensor under time reversal and parity transformation 
has the property \cite{pppol,Pick} 

\begin{equation}
W^{\mu\nu}({\s}, (-)) = W^{\nu\mu}(-{\s}, (+)) ,
\label{eq:pick}
\end{equation}
where ${\s}$ is the spin vector in the  ejectile rest frame, and the dependence 
on the final state boundary condition for incoming $(-)$ and outgoing $(+)$ 
scattered waves is shown. For nucleon knockout, the $(-)$ condition is 
appropriate. When the boundary conditions can be ignored, as in the plane-wave 
(PW) approximation, eq.~(\ref{eq:pick}) states that the symmetric part of 
$W^{\mu\nu}$ is independent of ${\s}$ and the antisymmetric part is proportional 
to ${\s}$. Therefore, when the hadron tensor is spin independent its 
antisymmetric part vanishes, while its symmetric part vanishes when it is spin 
dependent. Thus, in PW only $h^u_{11}$, $h^u_{1-1}$, 
$\bar{h}^u_{1-1}$, and $h'^{\mathrm N,L,S}_{11}$ survive and therefore 
$\sigma_0$, $\Sigma_{\frac {\pi} {2}}$, $\Sigma_{\frac {\pi} {4}}$, and 
$P^{\mathrm N,L,S}_{\mathrm c}$, while  $A_{\mathrm c}$ = 
$P^{\mathrm N,L,S}$ = $\Sigma^{\mathrm N,L,S}_{\frac {\pi} {2}}$ = 
$\Sigma^{\mathrm N,L,S}_{\frac {\pi} {4}}$ = 0.

If both coplanar kinematics and the PW approximation are considered, only 
$\sigma_0$, $\Sigma_{\frac {\pi} {2}}$ and $P^{\mathrm L,S}_{\mathrm c}$ 
survive. The results are summarized in table~1.

\section{The 
$^{16}$O($\vec{\gamma},\vec{\n}\p$)$^{14}$N reaction}
\label{sec:3}
In this section numerical results are presented for the 
$^{16}$O($\gamma,np$)$^{14}$N reaction. Calculations have been performed within 
the theoretical framework of ref. \cite{pn}. In this model it is possible to 
consider  transitions to different low-lying discrete states of the residual 
nucleus, that are expected to be strongly populated by direct knockout. Here, 
we present numerical examples for the transition to the 1$^+$ ground state of 
$^{14}$N, with $T=0$. This is the first state that can be experimentally 
separated with an energy resolution of a few MeV and is therefore of particular 
interest. 

\begin{figure}
\epsfysize=14.0cm
\begin{center}
\makebox[16.4cm][c]{\epsfbox{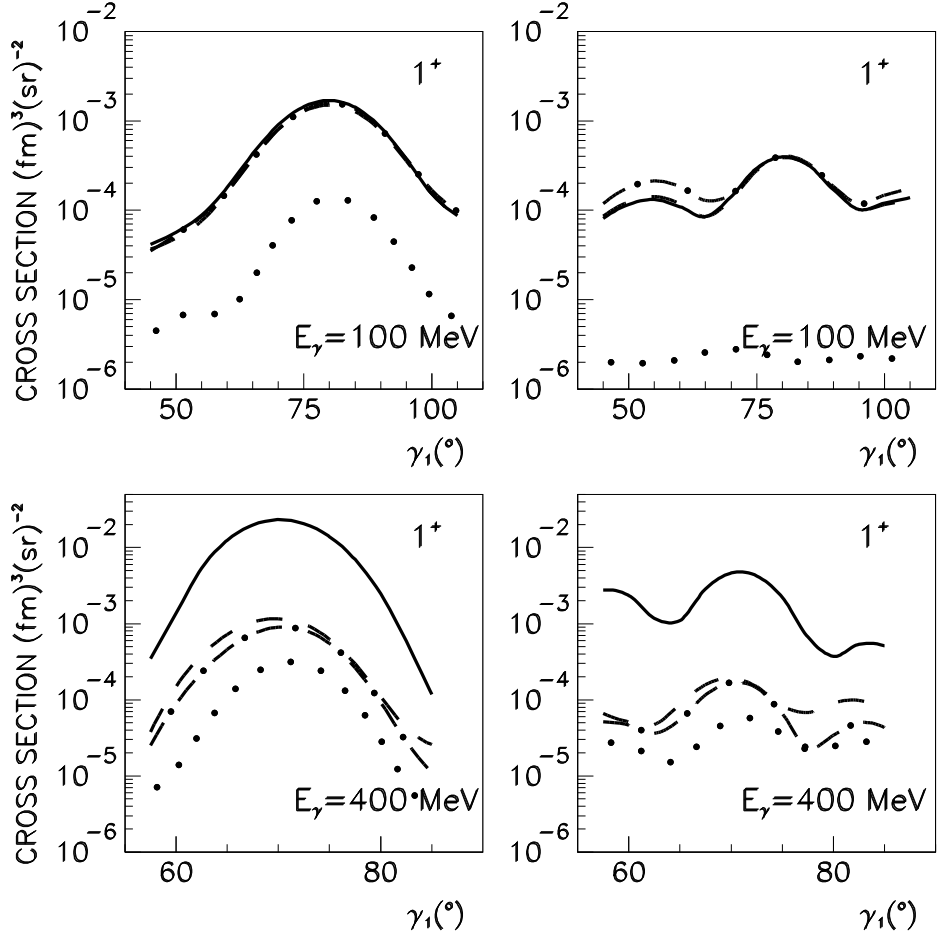}}
\end{center}
\caption[]{The differential cross section  of the
$^{16}$O($\gamma,np$)$^{14}$N$_{\mathrm{g.s.}}$ reaction as a function of the 
scattering angle $\gamma_1$ of the outgoing neutron in coplanar and symmetrical 
kinematics at $E_\gamma=100$ MeV (upper panels) and $E_\gamma=400$ MeV (lower 
panels). The two-nucleon overlap is calculated as in ref. \cite{pn} (left
panels) and as in ref. \cite{gnn} (right panels). The optical potential is taken 
from ref. \cite{Nad}. The separate contributions given by the one-body currents 
(dotted lines), the sum of the one-body and seagull currents (dot-dashed
lines), the sum of the one-body, seagull and pion-in-flight currents (dashed
lines) are displayed. The solid lines give the final cross sections, where also
the contribution of the $\Delta$ current is added.  
\label{fig:fig1}
}
\end{figure}

The results are very sensitive to the theoretical treatment of the two-nucleon 
overlap between the ground state of the target and the final state of the
residual nucleus. An example is shown in fig. \ref{fig:fig1}, where the 
calculated differential cross sections of the $^{16}$O($\gamma,np$) reaction 
are displayed in a coplanar and symmetrical kinematics at the incident photon 
energies of $E_\gamma = 100$ MeV and 400 MeV. In this kinematics the two 
nucleons are emitted in the scattering plane at equal energies and equal but 
opposite angles with respect to the beam direction. Then, for a fixed value of 
$E_\gamma$ and changing the value of the scattering angle it is possible to 
explore all values of the recoil ($p_{\mathrm{r}}$) or missing momentum 
($p_{2\mathrm{m}}$) distribution. One has 
\begin{equation}
\p_{2\mathrm{m}} = \p_{\mathrm{r}} =  \q -\p'_{1} - \p'_{2},  \label{eq:pm}
\end{equation}
where $\q$, $\p'_1$, and $\p'_2$ are the momenta of the incident photon and 
of the two outgoing nucleons, respectively. 

If relative and c.m. motions are factorized and final-state interactions are 
neglected \cite{Gottfried}, $\p_{\mathrm{r}}$ is opposite to the total 
momentum $\P$ of the nucleon pair in the target. Thus, the shape of the momentum 
distribution is driven by the c.m. orbital angular momentum of the knocked-out
pair. As this feature is not spoiled in an unfactorized approach, the 
symmetrical kinematics here considered is well suited to give information on 
the motion of the pair in different c.m. angular momentum states.

Calculations have been performed within the direct knockout framework of
refs. \cite{pn,gnn}, where the final-state wave function includes the
interaction of each one of the two outgoing nucleons with the residual nucleus
by means of a phenomenological optical potential. The nuclear current is the sum
of a one-body part, including convective and spin currents, and of a two-body
part, including terms corresponding to the lowest order diagrams with one-pion 
exchange, namely seagull, pion-in-flight and those with intermediate 
$\Delta$ isobar configurations. 

The results given by two different treatments of the two-nucleon overlap are 
compared in the left and right panels of fig. \ref{fig:fig1}. The cross sections displayed in the left panels are
obtained with the overlap integral used in \cite{pn} for the 
$^{16}$O($e,e'np$)$^{14}$N reaction. It includes the effects of short-range as 
well as tensor correlations, which are calculated within the framework of the 
coupled cluster method and with a correlation operator restricted to $1p-1h$ 
and $2p-2h$ excitations. The Argonne V14 potential \cite{v14} is employed as a 
model for a realistic nucleon-nucleon interaction. The explicit expression of
the two-nucleon overlap is given by an expansion over relative and c.m. wave
functions. The expansion coefficients, accounting for the global or long-range
structure of the specific nuclear states, are determined from a configuration
mixing calculation of the two-hole states in $^{16}$O, which can be coupled to 
the angular momentum and parity of the requested final state and are 
renormalized to account for the spectroscopic factors of the s.p. states. 
This is the best model which can be applied at present, as it contains all the
complications of a many-body calculation and the effects of a realistic
interaction.  

The results in the right panels are calculated with the simpler prescription 
used in ref. \cite{gnn}, \textit{i.e.} by the product of the pair function of 
the shell model, described for the $1^+$ ground state of $^{14}$N as a pure 
($p_{1/2}$)$^{-2}$ hole, and of a Jastrow type central and state independent 
correlation function \cite{GD}. With this simpler prescription the contribution
of the one-body currents and thus of correlations is much lower than with the
more refined model, where also tensor correlations are included. In both cases, 
however, and for both values of the photon energy, the calculated cross sections 
are dominated by two-body currents, \textit{i.e.} by the seagull current at 100 
MeV and by the $\Delta$ current at 400 MeV. The differences between the results 
shown in the left and right panels of the figure are anyhow large. They are in 
particular due to the presence in the model of ref. \cite{pn}, together with the 
($p_{1/2}$)$^{-2}$ component, with a strength 0.60, of an interference component 
($p_{1/2}p_{3/2}$)$^{-1}$ equal to $-0.45$~\cite{pn}. These differences, which 
are due only to the treatment of the two-nucleon overlap, clearly show that the 
shape and the size of the cross section are very sensitive to this treatment and 
that an accurate description of both aspects related to nuclear structure and 
correlations is needed to produce reliable numerical predictions. This result, 
that is well established for ($e,e'NN$) reactions 
\cite{Gerco,NIKHEF,Ronald,Rosner}, is 
therefore confirmed also for two-nucleon knockout induced by real photons. 

\begin{figure}
\epsfysize=14.0cm
\begin{center}
\makebox[16.4cm][c]{\epsfbox{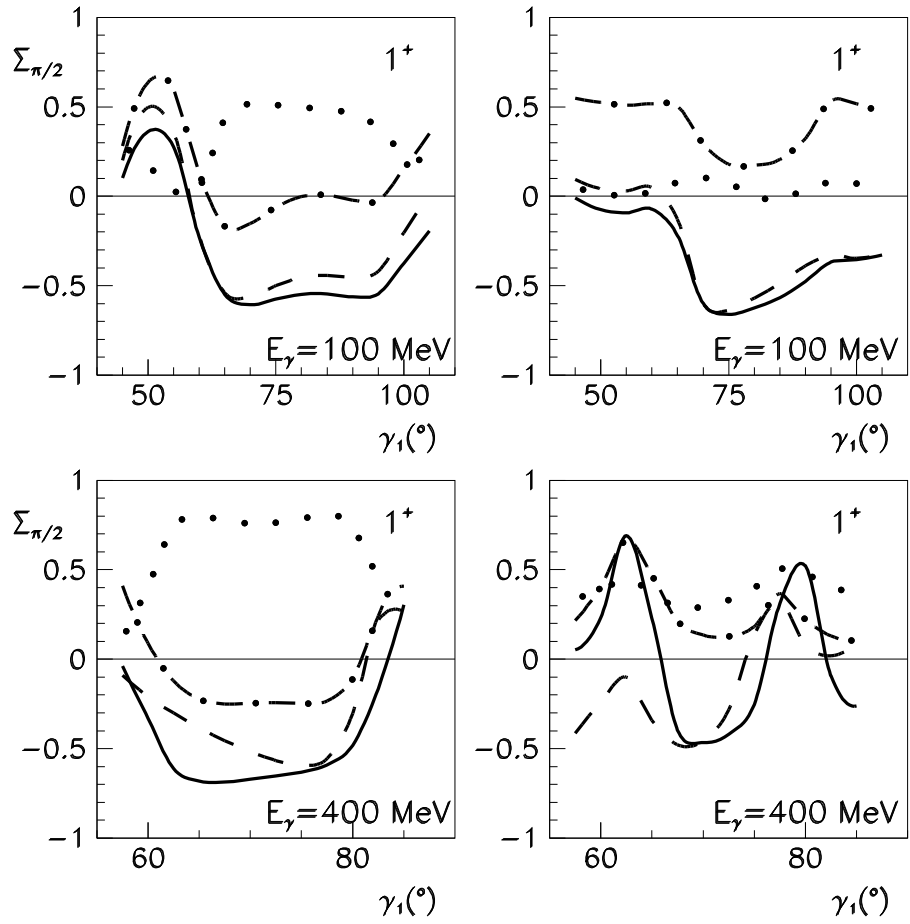}}
\end{center}
\caption[]{The photon asymmetry of the 
$^{16}$O($\gamma,np$)$^{14}$N$_{\mathrm{g.s.}}$ 
reaction in the same kinematics and conditions and with the same line convention 
as in fig. \ref{fig:fig1}.
\label{fig:fig2}
}
\end{figure}

The linear asymmetry $\Sigma_{\frac {\pi} {2}}$, calculated with the two 
different prescriptions for the overlap integral and in the same conditions and 
kinematics as in fig. \ref{fig:fig1}, is shown in fig. \ref{fig:fig2}. Large 
differences are given also in
this case by the two models, both  at $E_\gamma =$ 100 and 400 MeV. The 
contribution of the one-body current is generally small. This confirms that 
even when two-body currents are dominant the results can be very sensitive to 
nuclear structure and correlation effects. Moreover, as it was already observed 
in previous investigations \cite{gnn,Gent2}, the asymmetry appears particularly 
affected by the different terms of the two-body current and by their 
interference, in particular by the interference between the pion-in-flight and 
seagull currents. This contribution is dominant on the asymmetry at 
$E_\gamma = $ 100 MeV, while the role of pion-in-flight on the cross section is 
generally small. At $E_\gamma =$ 400 MeV  the $\Delta$ current gives the main
contribution, which, however, is not as dominant as in the cross section. 
Nonnegligible effects are given also 
by the other components of the two-body current.  

For the following calculations we have chosen a coplanar kinematics at 
$E_\gamma =$ 120 MeV. At this value of the photon energy, far from the peak of 
the $\Delta$  resonance, the contribution of the
two-body $\Delta$ current, that is an important but difficult ingredient
\cite{delta} of the theoretical model, is drastically reduced. 
However, the photon energy is large enough to obtain outgoing nucleons with a
reduced final state interaction. The contribution of meson exchange currents
 cannot be eliminated
and the chosen energy is a compromise between the different contributions.
The energy and the scattering angle of 
the outgoing neutron are fixed at 45 MeV and $45^{\mathrm{o}}$, respectively. 
This value of the scattering angle is taken in order to minimise the mutual 
interaction between the outgoing nucleons.
Different values of the recoil momentum can be obtained by varying the 
scattering angle of the outgoing proton. This kinematics appears within reach 
of available experimental facilities. 

\begin{figure}
\epsfysize=10.0cm
\begin{center}
\makebox[16.4cm][c]{\epsfbox{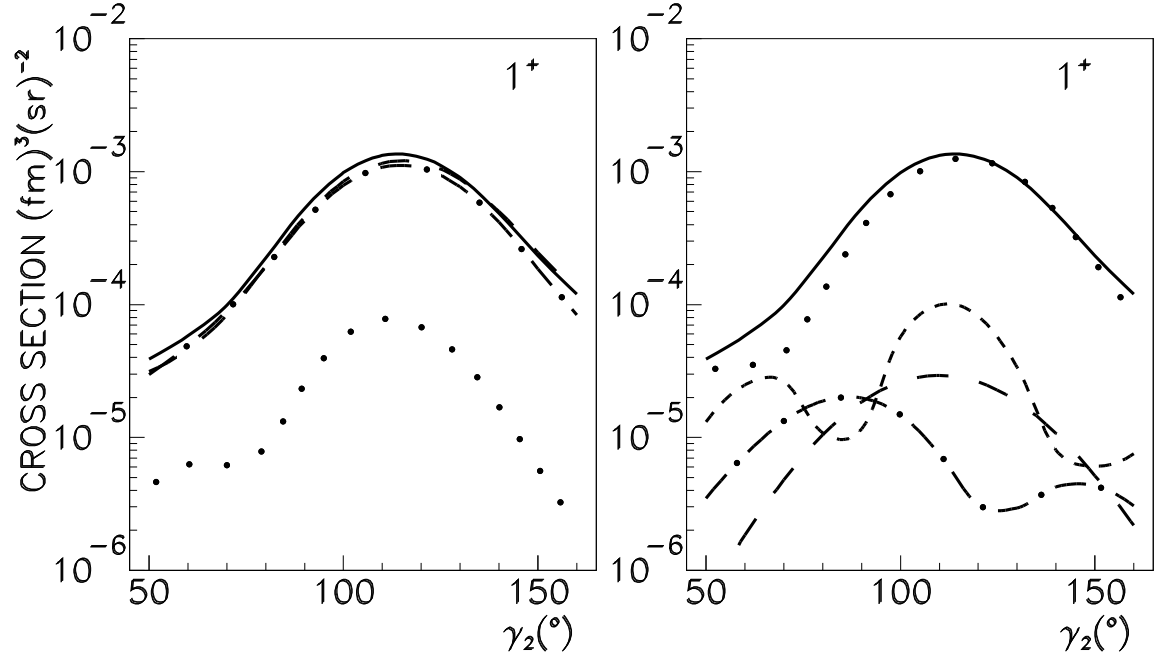}}
\end{center}
\caption[]{The differential cross section  of the
$^{16}$O($\gamma,np$)$^{14}$N$_{\mathrm{g.s.}}$ reaction as a function of the 
scattering angle $\gamma_2$ of the outgoing proton in a coplanar kinematics 
at $E_\gamma=120$ MeV, with an outgoing neutron energy 
$T_n = 45$ MeV and  $\gamma_1 =45^{\mathrm{o}}$.  The optical 
potential is taken from ref. \cite{Nad}. In the left panel the same line 
convention as  in fig. \ref{fig:fig1} is used. In the right panel separate 
contributions of different partial waves of relative motion are drawn: the 
dotted line is for $^3S_1$, the dot-dashed line is for $^1P_1$, the long-dashed 
line gives the $^3D_1$ component already present in the uncorrelated wave 
function and the short-dashed line gives the $^3D_1^T$ component due to tensor 
correlations. The solid line is the same as in the left panel. 
\label{fig:fig3}
}
\end{figure}

The calculated differential cross section is displayed in fig. \ref{fig:fig3}. 
In the left panel the results given by the one-body current and by the addition 
of the different terms of the two-body current are shown. At the considered 
value of the photon energy the pion seagull current gives the dominant 
contribution to the cross section. Only a minor role is played by the one-body 
current and by the pion-in flight and $\Delta$ currents. The shape of the angular 
distribution, however, is practically the same as the one given by the one-body 
current. 

In the considered kinematics and in the calculated angular range, the recoil 
momentum varies between $\simeq 90$ and 290 MeV/$c$, on both sides of the 
distribution.  The shape of the distribution is driven by the c.m. orbital 
angular momentum $L$ of the knocked-out pair. In our model the two-nucleon 
overlap function contains different components of relative and c.m. 
motion \cite{pn}. For the transition to the $1^+$ ground state the relative 
waves are: $^3S_1$ (the notation $^{2S+1}l_j$, for $l = S,P,D,$  is used here 
for the relative states), combined with a c.m. $L= 0$ and $L= 2$, $^1P_1$, 
combined with $L= 1$, and $^3D_1$. For this relative wave we have separated the 
component already present in the uncorrelated wave function, which is combined 
with $L= 0$, and the one produced by tensor correlations and not present in the 
uncorrelated wave function, which is combined with $L= 0$ and $L= 2$. As in 
ref. \cite{pn} we call these two terms $^3D_1$ and $^3D_1^{\mathrm T}$, 
respectively. The contribution of $^3D_1^{\mathrm T}$ emphasizes the role played 
in the calculations by tensor correlations, which are anyhow present also in the 
other components. 

The separate contributions of the different waves of relative motion are shown 
in the right panel of fig. \ref{fig:fig3}. The shape of the different curves is 
determined by the corresponding values of $L$. The shape of the final cross 
section is driven by the component which gives the major contribution, 
\textit{i.e.} by $^3S_1$ in the considered case. Only small effects are given by 
the other partial waves. The component $^3D_1^{\mathrm T}$, in particular its 
part with $L = 2$, is nonnegligible at low values of the scattering angle, 
which correspond to larger values of the momentum. The cross section is thus 
basically a combination of states with $L= 0$ and $L= 2$. The contribution of 
the state with $L= 1$ for $^1P_1$ is very small. The main role is played by the 
state with $L = 0$ for $^3S_1$, which gives the $s$-wave shape in the figure. 
The contribution of states with $L= 2$ for $^3S_1$ and $^3D_1^{\mathrm T}$ 
becomes important at large values of the recoil momentum, where the contribution 
of the state with $L = 0$ becomes much lower. 

It has been shown in sect.~\ref{sec:2} that when the incident photon is 
polarized and/or the polarization of the outgoing nucleon is detected, 7 
polarization observables survive in coplanar kinematics: the asymmetry 
$\Sigma_{\frac {\pi} {2}}$, that can be measured in a reaction with a linearly 
polarized photon and where the nucleon recoil polarization is not detected, the 
component $P^{\mathrm N}$ of the outgoing nucleon polarization, that can be 
measured in a reaction with an unpolarized incident photon, the components 
$P^{\mathrm{L,S}}_{\mathrm c}$ of the polarization transfer coefficient, that 
can be measured in a reaction with a circularly polarized photon, and the 
components $\Sigma^{\mathrm N}_{\frac {\pi} {2}}$ and 
$\Sigma^{\mathrm{L,S}}_{\frac {\pi} {4}}$ of the polarization asymmetries, that
can be measured in a reaction with a linearly polarized incident photon. 

\begin{figure}
\epsfysize=14.0cm
\begin{center}
\makebox[16.4cm][c]{\epsfbox{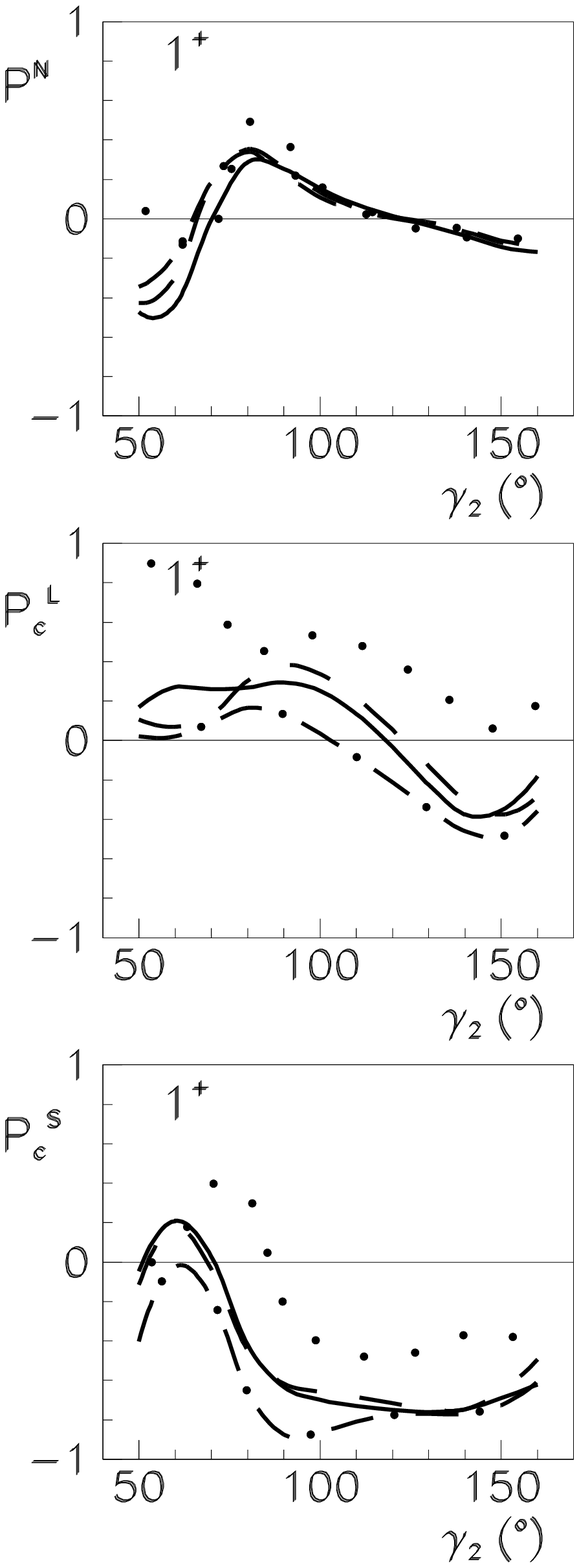}}
\end{center}
\caption[]{The polarization observables $P^{\mathrm N}$, 
$P_{\mathrm{c}}^{\mathrm L}$, and $P_{\mathrm{c}}^{\mathrm S}$ for the same 
reaction and in the same conditions and kinematics as in fig. \ref{fig:fig3}. 
Line convention as in the left panel of fig. \ref{fig:fig3}.
\label{fig:fig4}
}
\end{figure}

\begin{figure}
\epsfysize=14.0cm
\begin{center}
\makebox[16.4cm][c]{\epsfbox{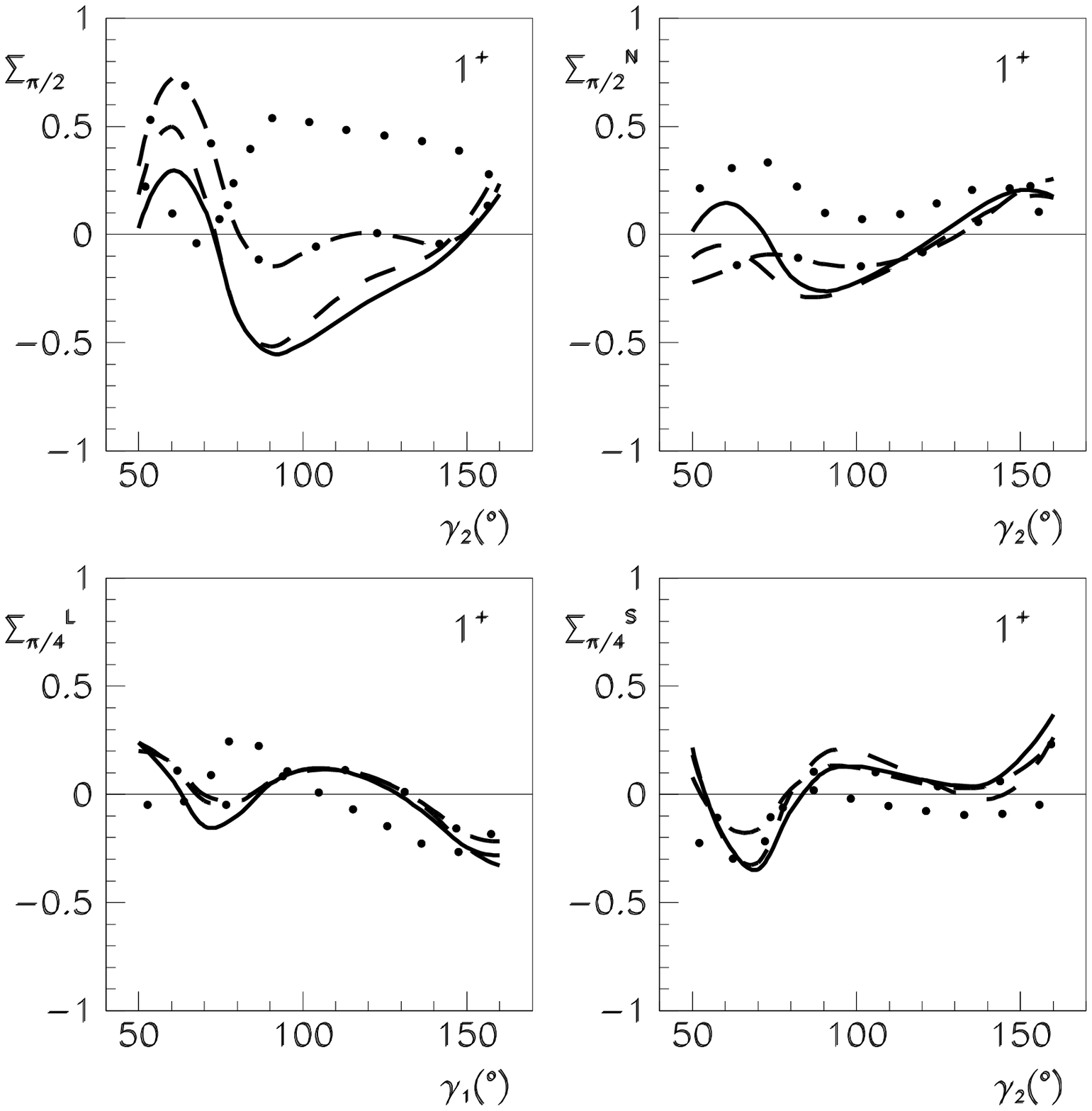}}
\end{center}
\caption[]{The polarization observables $\Sigma_{\pi/2}$ 
$\Sigma_{\pi/2}^{\mathrm N}$, $\Sigma_{\pi/4}^{\mathrm L}$, and 
$\Sigma_{\pi/4}^{\mathrm S}$ for the same reaction, in the same conditions and 
kinematics and with the same line convention as in the left panel of 
fig. \ref{fig:fig3}. 
\label{fig:fig5}
}
\end{figure}

The polarization observables calculated with our model and in the coplanar
kinematics of fig. \ref{fig:fig3} are shown in figs. \ref{fig:fig4}
and \ref{fig:fig5}. The results indicate that all the observables are sizable. 
They are sensitive to all the different terms of the nuclear current and also 
to their interference. The main contribution is generally given by the seagull 
current, which, however, is not as dominant as in the cross section. A large 
effect is given by the interference between the seagull and pion-in-flight 
terms, in particular for the asymmetry $\Sigma_{\frac {\pi} {2}}$, where this 
effect gives the main contribution. The role of the one-body current is not very 
important in general, but it can be meaningful in particular situations, for 
instance on $P^{\mathrm N}$. At the considered value of the photon energy, 
$E_\gamma = 120$ MeV, a small although nonnegligible effect is generally given 
by the $\Delta$ current. 

We note that $\Sigma^{\mathrm N}_{\frac {\pi} {2}}$ and 
$\Sigma^{\mathrm{L,S}}_{\frac {\pi} {4}}$ vanish in the PW approximation, when
final-state interactions (FSI) are neglected. Indeed in fig. \ref{fig:fig5} 
these quantities turn out to be smaller than the other observables which are 
present also in PW. Since they are produced by FSI, one might expect that they 
are sensitive to their treatment.  

The results depend on kinematics and on the final state that is considered. The
numerical example in figs. \ref{fig:fig3}-\ref{fig:fig5} indicates that cross 
sections and polarization observables show a different sensitivity to the
different terms of the nuclear current. This can be understood if one considers
that different structure functions contribute to the different quantities. 
Moreover, since the polarization observables can be expressed in terms of 
ratios between the various structure functions and $h^u_{11}$, that 
gives the unpolarized cross section, one can expect that they are able to 
emphasize specific effects that can be smoothed out in the cross section. Thus, 
a combined experimental determination of cross sections and polarization 
observables would give a complete information on the reaction process. It would 
impose constraints to the different theoretical ingredients and result in a 
stringent test of theoretical models. In figs. \ref{fig:fig3}-\ref{fig:fig5} we 
have shown the sensitivity to the different terms of the nuclear current. In 
figs. \ref{fig:fig1} and \ref{fig:fig2}, in different kinematics, it has been 
shown that the cross section and the photon asymmetry $\Sigma_{\frac {\pi} {2}}$ 
exhibit a strong and different sensitivity to the theoretical treatment of the 
two-nucleon overlap. 

\begin{figure}
\epsfysize=10.0cm
\begin{center}
\makebox[16.4cm][c]{\epsfbox{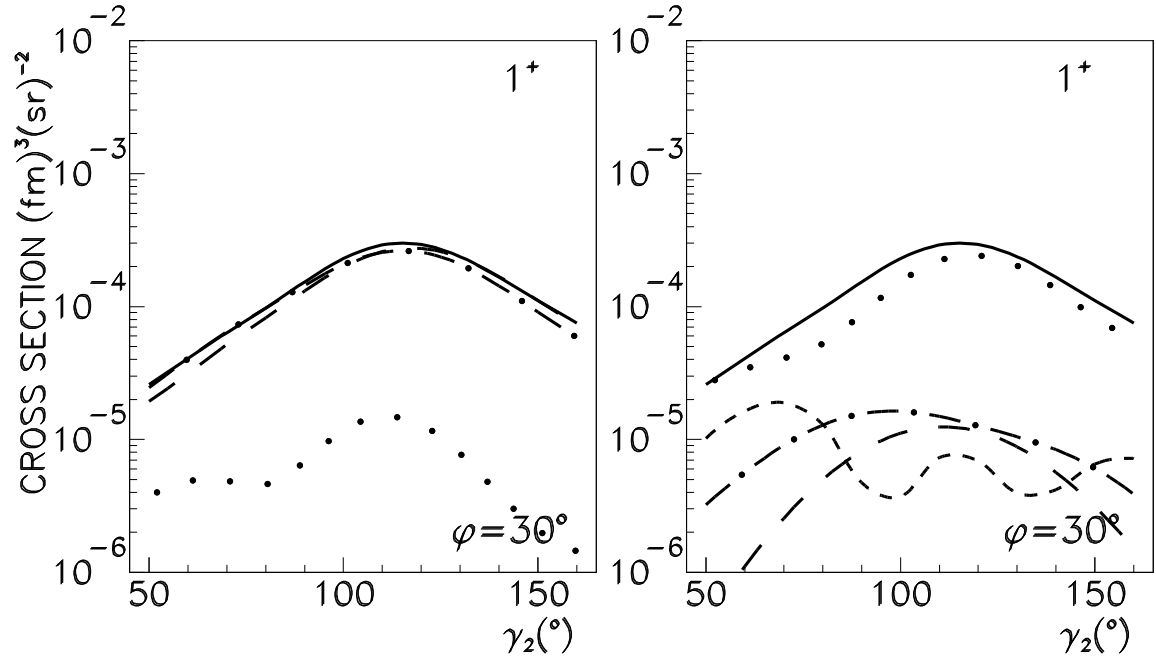}}
\end{center}
\caption[]{The differential cross section of the
$^{16}$O($\gamma,np$)$^{14}$N$_{\mathrm{g.s.}}$ reaction as a function of the 
scattering angle $\gamma_2$ of the outgoing proton in an out-of-plane kinematics 
at $E_\gamma=120$ MeV, with $T_n = 45$ MeV, 
$\gamma_1 =45^{\mathrm{o}}$ and the azimuthal angle of the outgoing proton
$\phi = 30^{\mathrm{o}}$. The optical potential is taken from ref. \cite{Nad}. 
Line convention in the left and right panels as in fig. \ref{fig:fig3}.
\label{fig:fig6}
}
\end{figure}

\begin{figure}
\epsfysize=14.0cm
\begin{center}
\makebox[16.4cm][c]{\epsfbox{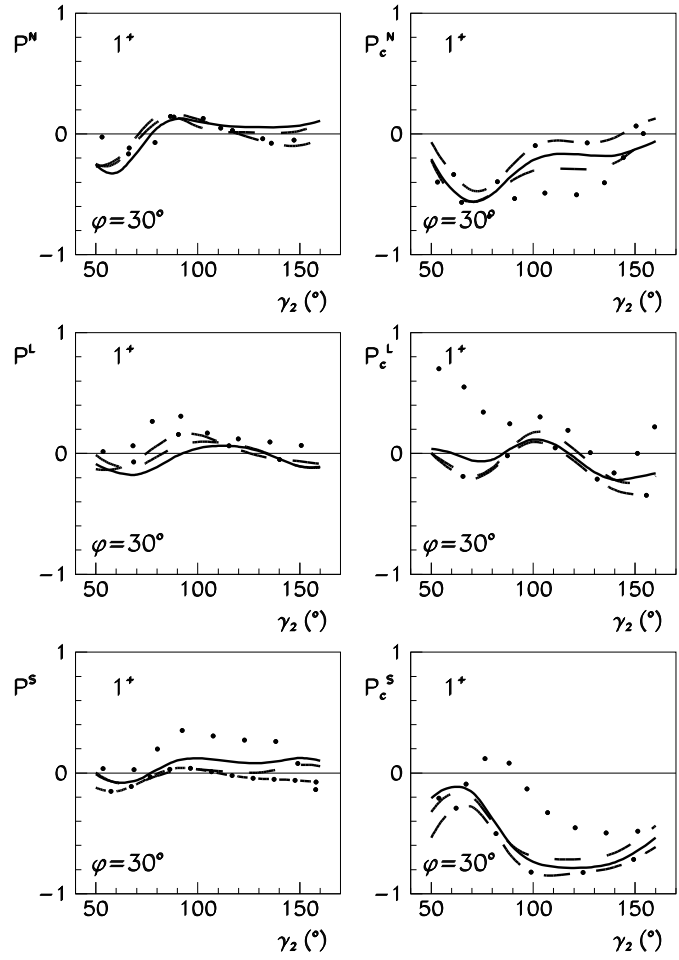}}
\end{center}
\caption[]{The polarization observables $P^{\mathrm N}$,
$P_{\mathrm{c}}^{\mathrm N}$, $P^{\mathrm L}$, $P_{\mathrm{c}}^{\mathrm L}$, 
$P^{\mathrm S}$, and $P_{\mathrm{c}}^{\mathrm S}$ for the same reaction and in 
the same conditions and kinematics as in fig. \ref{fig:fig6}. Line convention 
as in the left panel of fig. \ref{fig:fig3}.
\label{fig:fig7}
}
\end{figure}

\begin{figure}
\epsfysize=14.0cm
\begin{center}
\makebox[16.4cm][c]{\epsfbox{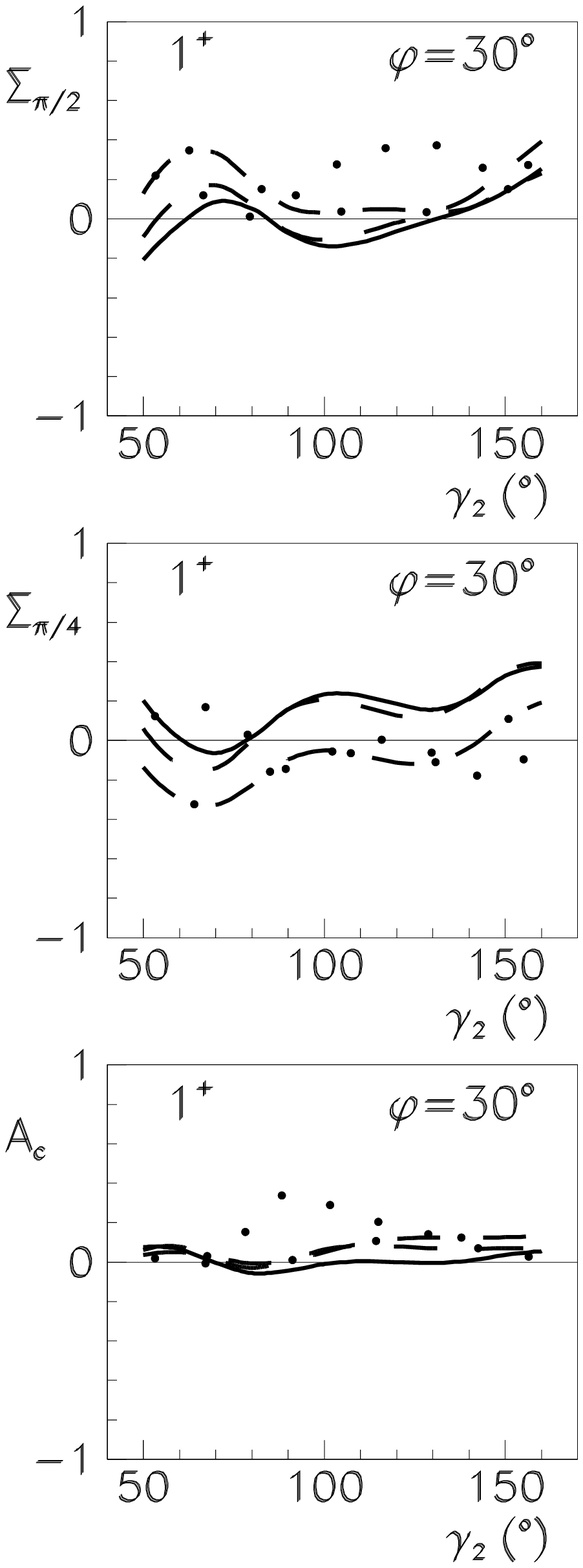}}
\end{center}
\caption[]{The polarization observables $\Sigma_{\pi/2}$, $\Sigma_{\pi/4}$,
and $A_{\mathrm{c}}$ for the same reaction and in the same conditions and 
kinematics as in fig. \ref{fig:fig6}. Line convention as in the left panel of 
fig. \ref{fig:fig3}.
\label{fig:fig8}
}
\end{figure}

\begin{figure}
\epsfysize=14.0cm
\begin{center}
\makebox[16.4cm][c]{\epsfbox{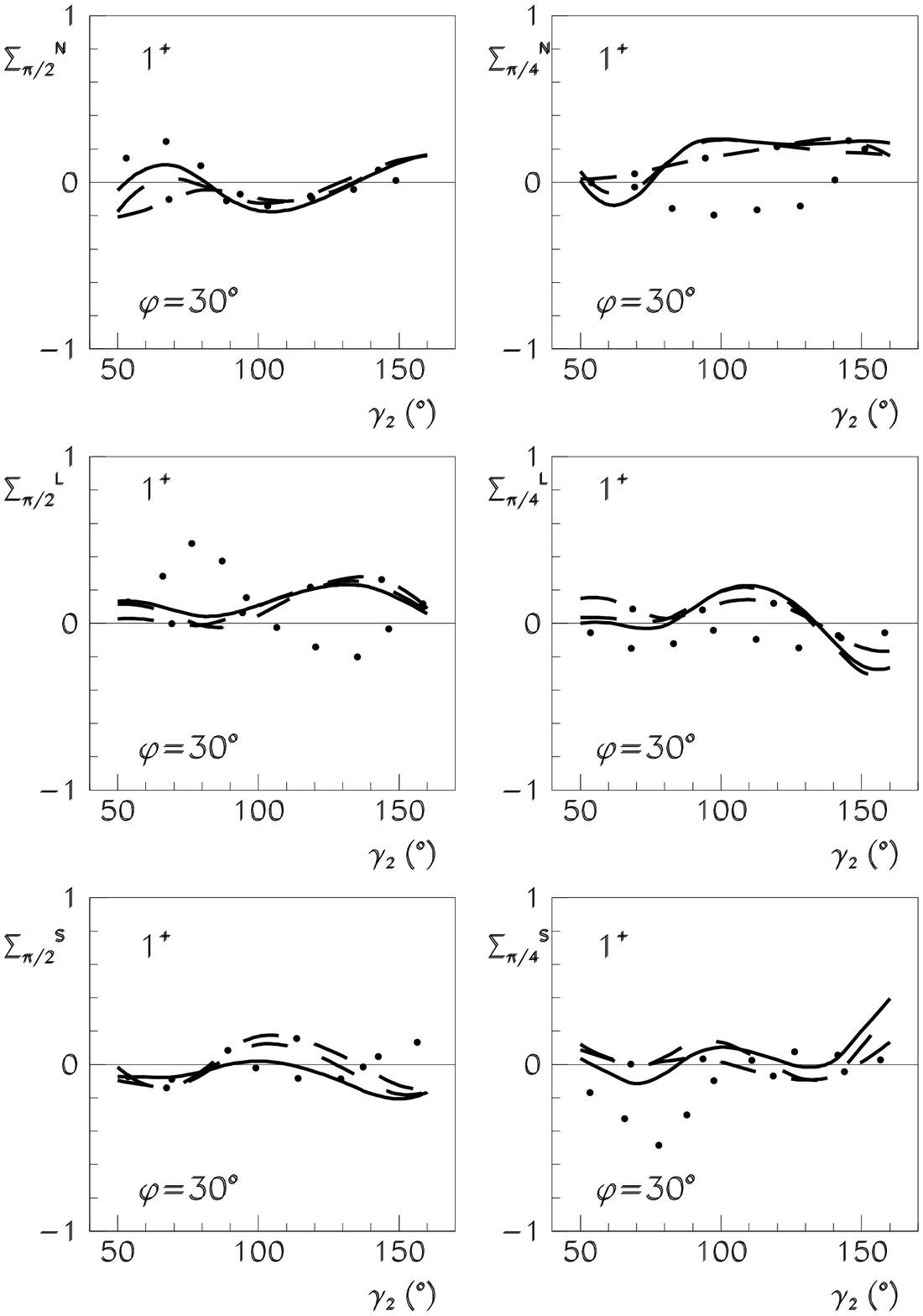}}
\end{center}
\caption[]{The polarization observables $\Sigma_{\pi/2}^{\mathrm N}$, 
$\Sigma_{\pi/4}^{\mathrm N}$, $\Sigma_{\pi/2}^{\mathrm L}$, 
$\Sigma_{\pi/4}^{\mathrm L}$, $\Sigma_{\pi/2}^{\mathrm S}$, and 
$\Sigma_{\pi/4}^{\mathrm S}$ for the same reaction and in the same conditions 
and kinematics as in fig. \ref{fig:fig6}. Line convention as in the left panel
of fig. \ref{fig:fig3}.
\label{fig:fig9}
}
\end{figure}

A complete study of all the 16 observables defined in sect.~\ref{sec:2} 
requires an out-of-plane kinematics. A numerical example is shown in 
figs. \ref{fig:fig6}-\ref{fig:fig9} in a situation where the azimuthal angle 
$\phi$ of the outgoing nucleon whose polarization is not considered (the proton 
in our case) is $30^{\mathrm{o}}$. The other kinematical variables are taken as 
in the coplanar kinematics of fig. \ref{fig:fig3}. 

The differential cross section in fig. \ref{fig:fig6} confirms the dominant 
role of the seagull current and of the $^3S_1$ component. The size, however, is 
an order of magnitude lower than in the peak region of the coplanar kinematics, 
while the shape indicates a less pronounced peak and a larger contribution of 
the components with $L = 2$. The main reason of the differences can be 
attributed to a kinematic effect. Different values of the recoil momentum are 
obtained for the same angle $\gamma_2$ in the two kinematics. When the outgoing 
nucleon is taken out of the plane, values lower than $160$ MeV/$c$ are forbidden 
and the region between 90 and 160 MeV/$c$, where the cross section in the 
coplanar kinematics has the maximum, is cut. This explains the different size 
and shape of the results in  figs. \ref{fig:fig3} and \ref{fig:fig6}. This 
kinematic effect gives also the main difference between the calculated 
polarization observables which are present both in coplanar and out-of-plane 
kinematics. The other observables, which are nonvanishing only in out-of-plane 
kinematics, are sizable. In the considered situation the main contribution is 
generally given by the seagull current, but also the other terms, in particular 
the interference between seagull and pion-in-flight, can be meaningful in some 
observables.
\section{The 
$^{16}$O($\vec{\gamma},\vec{\p}\p$)$^{14}$C reaction}
\label{sec:4}
In this section numerical results are presented for the 
$^{16}$O($\gamma,pp$)$^{14}$C reaction. Calculations have been performed within 
the theoretical framework of ref. \cite{sf}. The model is basically the same
used for the $^{16}$O($\gamma,np$)$^{14}$N reaction where the nuclear current is
adapted to a $pp$ pair. Thus, in the two-body current the charge-exchange terms 
vanish and only a part of the $\Delta$ current contributes. However, a different 
treatment of the two-nucleon overlap function is used, based on the results 
of the calculation of the two-proton spectral function of 
$^{16}$O \cite{Geurts,sf}. The two-nucleon overlaps for transitions to the 
lowest-lying discrete final states of the residual nucleus are obtained from a 
two-step procedure, where long-range and short-range correlations are treated in 
a separate but consistent way. The calculation of long-range correlations is 
performed in a shell-model space large enough to incorporate the corresponding 
collective features which influence the pair removal amplitudes. The 
single-particle propagators used for this dressed Random Phase Approximation 
description of the two-particle propagator also include the effect of both 
long-range and short-range correlations. In the second step that part of the 
pair removal amplitudes which describes the relative motion of the pair is 
supplemented by defect functions, accounting for SRC and obtained from the same 
G-matrix which is also used as the effective interaction in the RPA calculation. 
The explicit expression of the overlap function is given in terms of a sum of 
products of relative and c.m. wave functions. Different components contribute to
different transitions. Here, we present numerical results for the transition to 
the 0$^+$ ground state of $^{14}$C, where the components are: $^1S_0$, which is 
combined with $L= 0$, and $^3P_1$, which is combined with $L= 1$. The ground 
state has been already separated in recent high-resolution experiments for the 
$^{16}$O($e,e'pp$)$^{14}$C reaction \cite{Gerco,NIKHEF,Ronald,Rosner}. The 
experimental cross sections are well reproduced by the results of our model and 
clear evidence for SRC has been obtained in the comparison. Thus, it seems 
interesting to investigate the same final state also for the ($\gamma,pp$) 
reaction. 

\begin{figure}
\epsfysize=10.0cm
\begin{center}
\makebox[16.4cm][c]{\epsfbox{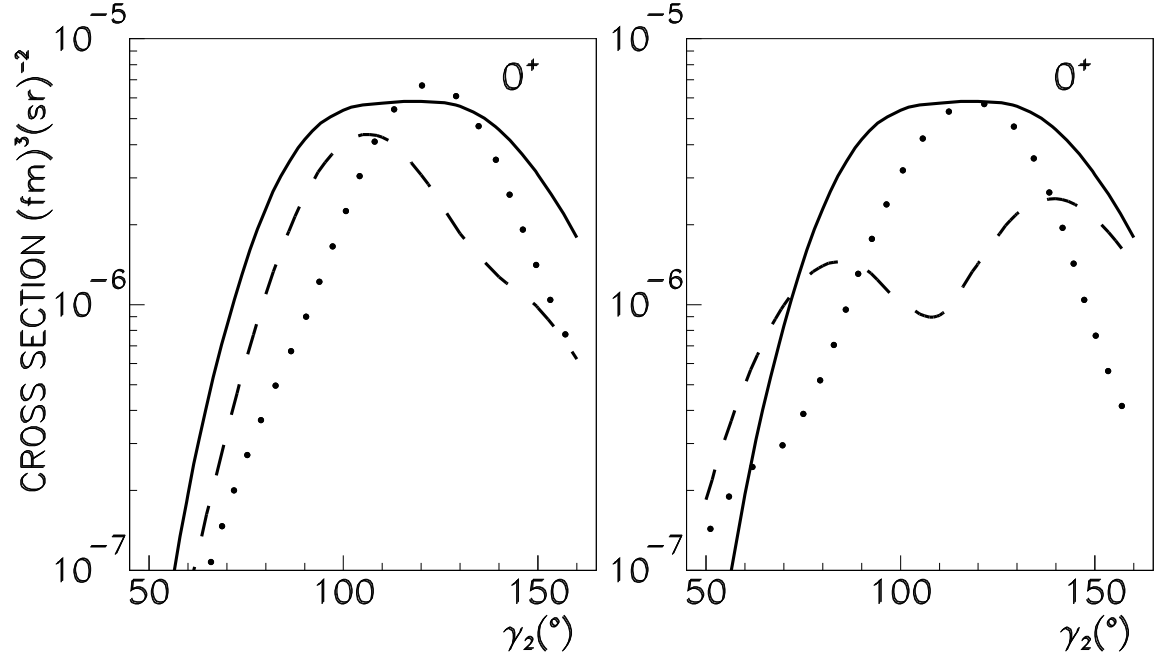}}
\end{center}
\caption[]{The differential cross section  of the
$^{16}$O($\gamma,pp$)$^{14}$C$_{\mathrm{g.s.}}$ reaction as a function of the 
scattering angle $\gamma_2$ in the same coplanar kinematics as in 
fig. \ref{fig:fig3}. The defect functions for the Bonn-A potential and the 
optical potential of ref. \cite{Nad} are used. The solid line gives the sum of 
the one-body and the two-body $\Delta$-current. In the left panel the separate 
contributions given by the one-body currents (dotted line) and by the $\Delta$ 
current (dashed line) are drawn. In the right panel separate contributions of 
different partial waves of relative motion are drawn: the dotted line is for
$^1S_0$, the dashed line is for $^3P_1$. 
\label{fig:fig10}
}
\end{figure}

The differential cross section of the $^{16}$O($\gamma,pp$)$^{14}$C reaction, 
calculated in the same coplanar kinematics considered in sect.~\ref{sec:3}, is 
displayed in fig. \ref{fig:fig10}. The shape of the distribution is determined 
by the combination of the two components with $L= 0$ and $L= 1$. It is clearly 
shown in the right panel of the figure that $^1S_0$ with its $s$-wave c.m. 
component prevails for angles corresponding to low values of the recoil 
momentum, where the $p$-wave has the minimum, while the contribution of $^3P_1$ 
knockout becomes important al large values of $p_{\mathrm r}$, where the 
$s$-wave decreases. 

It was established in previous investigations that SRC and two-body current play 
a different role in different relative states \cite{Gerco,Ronald,sf,pppol}. 
While SRC are quite strong for the $^1S_0$ state and much weaker for the $^3P_j$ 
states, the contribution of the two-body $\Delta$ current is strongly reduced 
in $^1S_0$ $pp$ knockout \cite{gnn,delta1} and generally dominant in $^3P_j$ 
knockout. Therefore, $^1S_0$ $pp$ knockout is generally dominated by the 
contribution of the one-body current, and thus by SRC, while $^3P_j$ $pp$ 
knockout is generally dominated by the $\Delta$ current. The selectivity  of 
the reaction to different states was exploited in ($e,e'pp$), where it was 
possible to envisage situations where either contribution is dominant and can 
thus be disentangled. The result, however, depends on kinematics and on the 
reaction that is considered. Thus, it cannot be directly applied to 
($\gamma,pp$) reactions, where only the transverse components of the nuclear 
current contribute and the longitudinal component, where the effects of SRC show 
up more strongly, is absent. On the other hand, in the considered kinematics, 
where $E_\gamma = 120$ MeV, the contribution of the $\Delta$ current should not 
be emphasized, as it is also demonstrated by the results of sect.~\ref{sec:3} 
for the ($\gamma,np$) reaction.  

In the present calculation the $\Delta$ current gives indeed the major 
contribution to $^3P_1$ knockout, but a sizable effect is also given by its 
interference with the one-body current. The one-body current is dominant in 
$^1S_0$ knockout. In the region of the maximum, where $^3P_1$ has the 
minimum, the contribution of the $\Delta$ current is larger in  $^1S_0$ than in 
$^3P_1$. This explains the result in the left panel of fig. \ref{fig:fig10}. 
Most of the contribution of the one-body current is given by $^1S_0$ knockout. 
This determines the $s$-wave shape of the angular distribution. The result for 
the $\Delta$ current is due to the combined effect of $^3P_1$, which gives the 
main contribution at large values of $p_{\mathrm r}$, and of $^1S_0$, which 
enhances the cross section with the $\Delta$ current at low values of 
$p_{\mathrm r}$. In the final cross section both processes due to SRC and to the 
two-body current are important: SRC are of particular relevance at larger values 
of $\gamma_2$ and the $\Delta$ at lower values of $\gamma_2$. 

\begin{figure}
\epsfysize=14.0cm
\begin{center}
\makebox[16.4cm][c]{\epsfbox{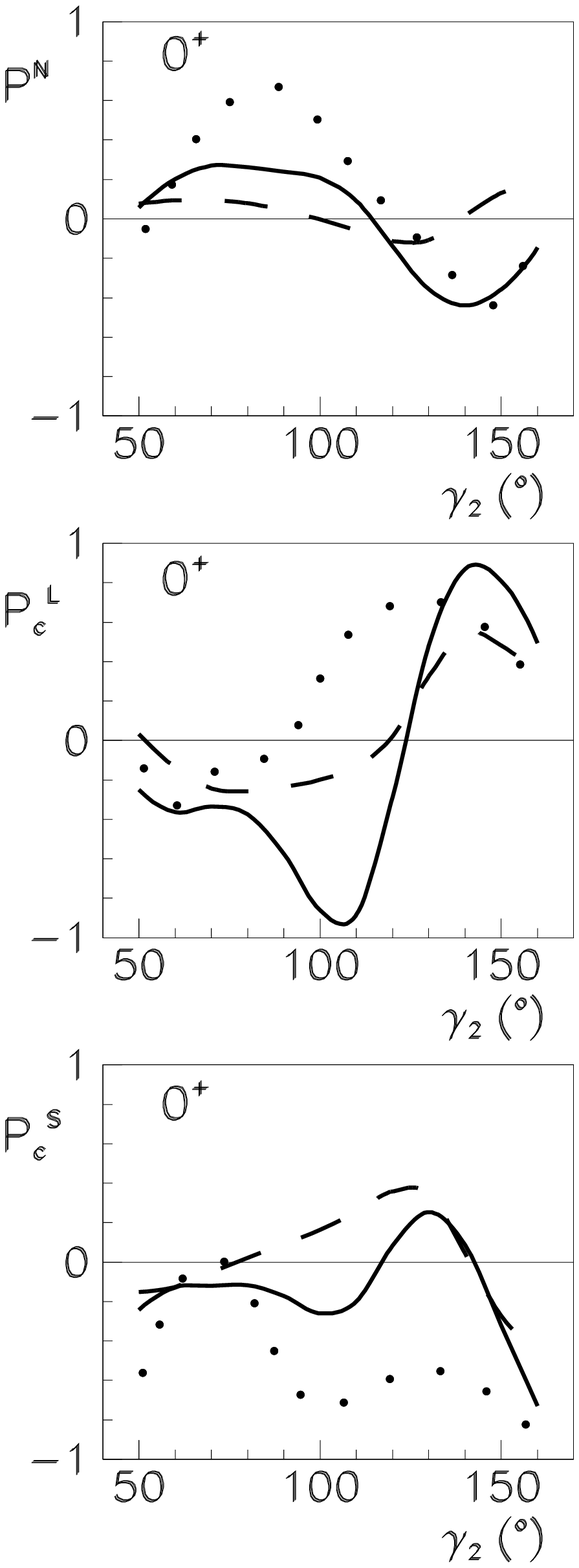}}
\end{center}
\caption[]{The polarization observables $P^{\mathrm N}$, 
$P_{\mathrm{c}}^{\mathrm L}$, and $P_{\mathrm{c}}^{\mathrm S}$ for the same 
reaction and in the same conditions and kinematics as in fig. \ref{fig:fig10}. 
Line convention as in the left panel of fig. \ref{fig:fig10}.
\label{fig:fig11}
}
\end{figure}

\begin{figure}
\epsfysize=14.0cm
\begin{center}
\makebox[16.4cm][c]{\epsfbox{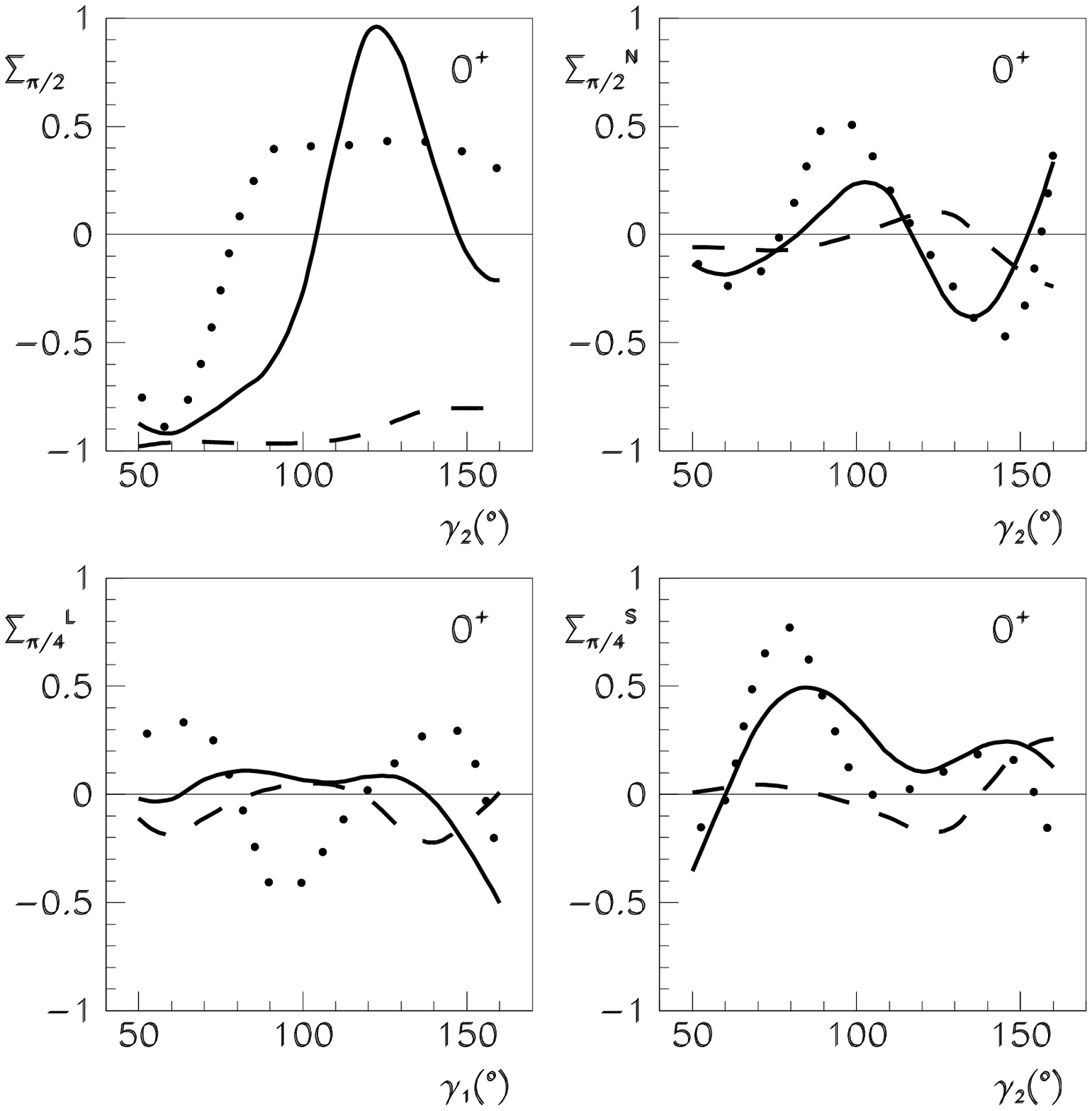}}
\end{center}
\caption[]{The polarization observables $\Sigma_{\pi/2}$ 
$\Sigma_{\pi/2}^{\mathrm N}$, $\Sigma_{\pi/4}^{\mathrm L}$, and 
$\Sigma_{\pi/4}^{\mathrm S}$ for the same reaction, in the same conditions and 
kinematics and with the same line convention as in the left panel of 
fig. \ref{fig:fig10}. 
\label{fig:fig12}
}
\end{figure}

The polarization observables are displayed in figs. \ref{fig:fig11} 
and \ref{fig:fig12}. All the calculated quantities are sizable. Both
contributions of the one-body and of the two-body current are important, even
though the relevance of the two reaction mechanisms can be different for
different observables. For instance, $\Sigma^{\mathrm N}_{\frac {\pi} {2}}$ is
basically given by the one-body current, while for other observables the final 
result, given by the sum of the one-body and the two-body current, is very 
different from the two separate contributions. Thus, a combined measurement of 
cross sections and polarization observables would be able to disentangle the two 
reaction processes and to test the ingredients and the approximations of a 
theoretical model.      

\begin{figure}
\epsfysize=10.0cm
\begin{center}
\makebox[16.4cm][c]{\epsfbox{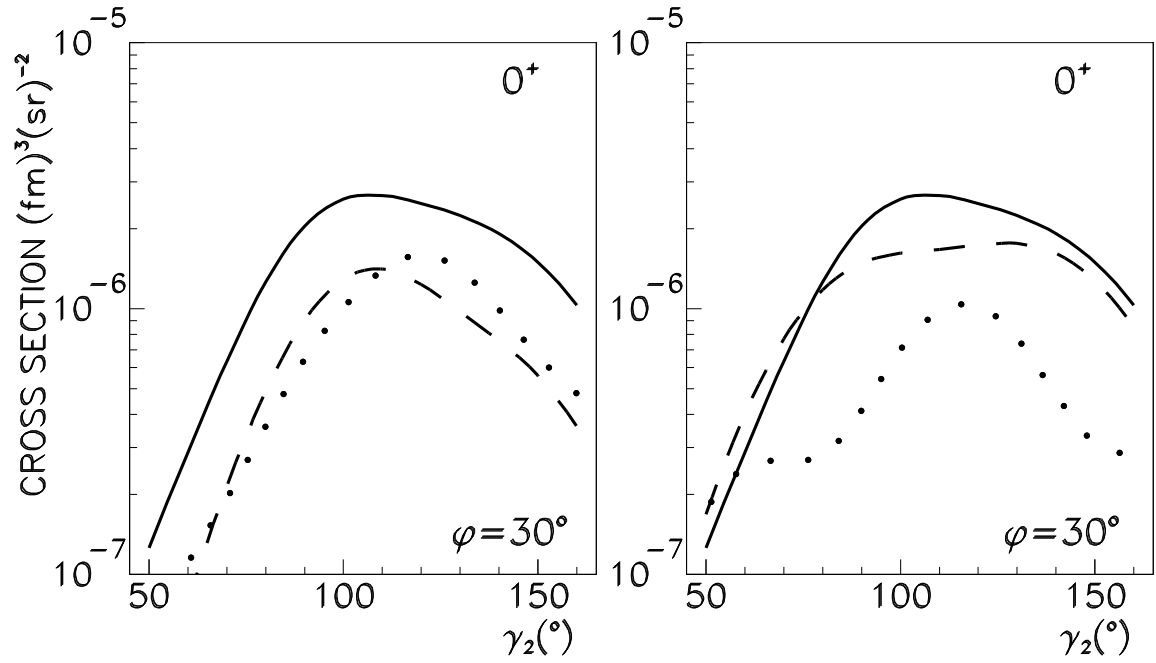}}
\end{center}
\caption[]{The differential cross section of the
$^{16}$O($\gamma,pp$)$^{14}$C$_{\mathrm{g.s.}}$ reaction as a function of the 
scattering angle $\gamma_2$ in the same out-of-plane kinematics as in 
fig. \ref{fig:fig6}. Defect functions, optical potential and line convention as 
in fig. \ref{fig:fig10}.
\label{fig:fig13}
}
\end{figure}

\begin{figure}
\epsfysize=14.0cm
\begin{center}
\makebox[16.4cm][c]{\epsfbox{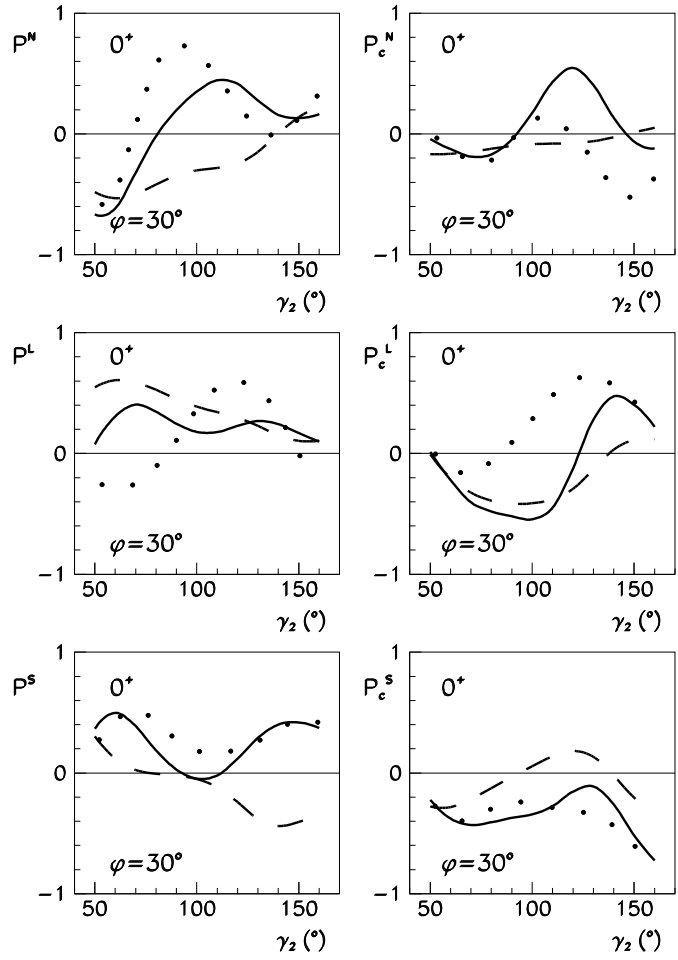}}
\end{center}
\caption[]{The polarization observables $P^{\mathrm N}$,
$P_{\mathrm{c}}^{\mathrm N}$, $P^{\mathrm L}$, $P_{\mathrm{c}}^{\mathrm L}$, 
$P^{\mathrm S}$, and $P_{\mathrm{c}}^{\mathrm S}$ for the same reaction and in 
the same conditions and kinematics as in fig. \ref{fig:fig13}. Line 
convention as in the left panel of fig. \ref{fig:fig10}.
\label{fig:fig14}
}
\end{figure}

\begin{figure}
\epsfysize=14.0cm
\begin{center}
\makebox[16.4cm][c]{\epsfbox{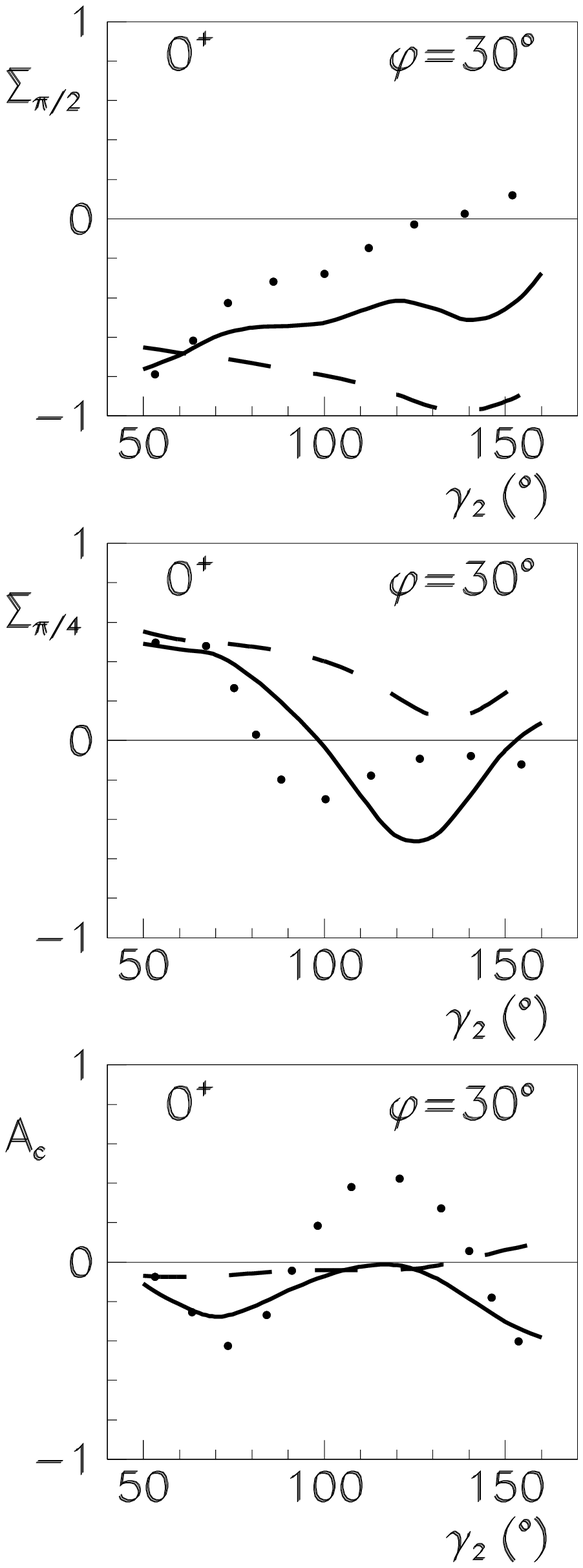}}
\end{center}
\caption[]{The polarization observables $\Sigma_{\pi/2}$, $\Sigma_{\pi/4}$,
and $A_{\mathrm{c}}$ for the same reaction and in the same conditions and 
kinematics as in fig. \ref{fig:fig13}. Line convention 
as in the left panel of fig. \ref{fig:fig10}.
\label{fig:fig15}
}
\end{figure}

\begin{figure}
\epsfysize=14.0cm
\begin{center}
\makebox[16.4cm][c]{\epsfbox{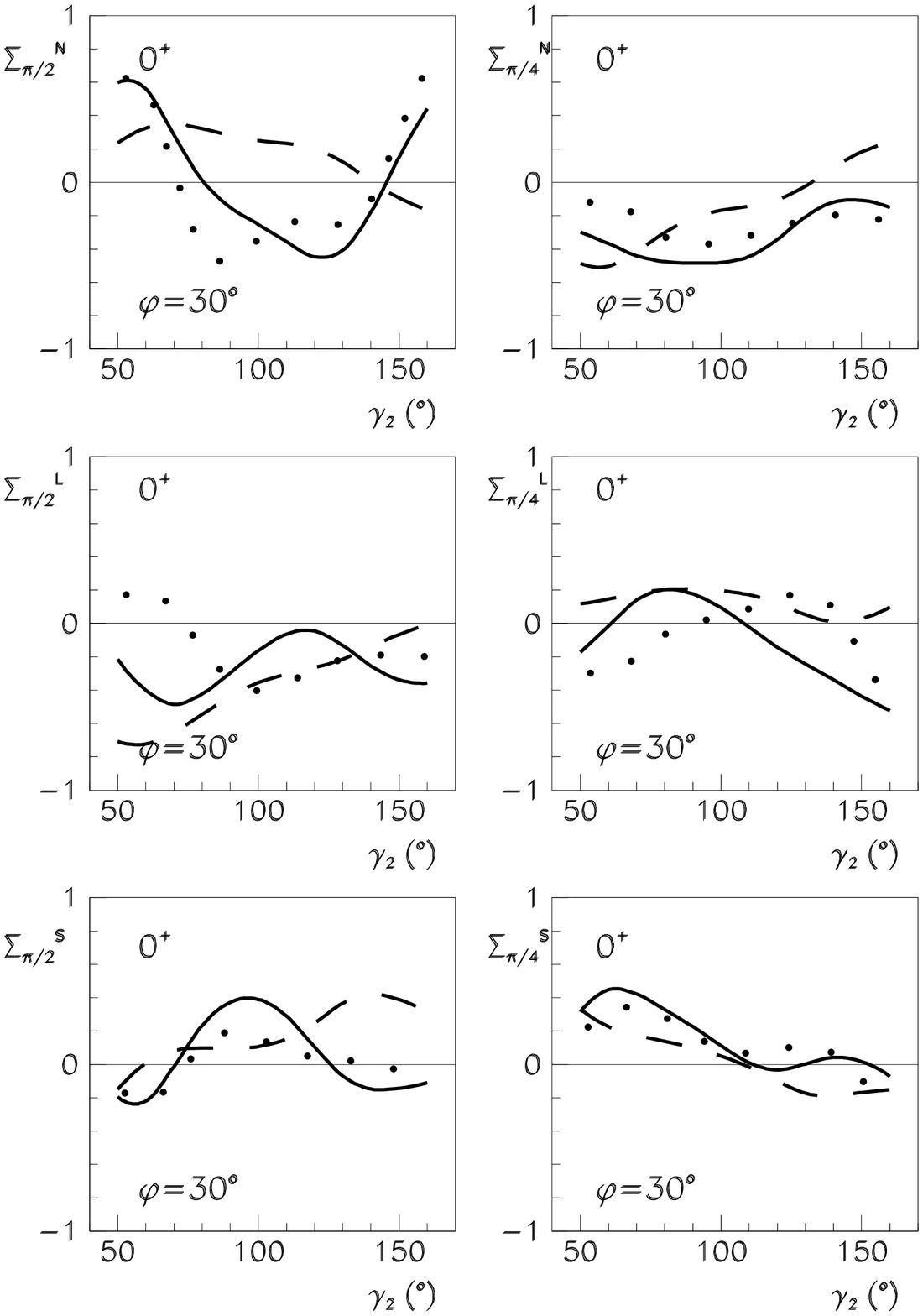}}
\end{center}
\caption[]{The polarization observables $\Sigma_{\pi/2}^{\mathrm N}$, 
$\Sigma_{\pi/4}^{\mathrm N}$, $\Sigma_{\pi/2}^{\mathrm L}$, 
$\Sigma_{\pi/4}^{\mathrm L}$, $\Sigma_{\pi/2}^{\mathrm S}$, and 
$\Sigma_{\pi/4}^{\mathrm S}$ for the same reaction and in the same conditions 
and kinematics as in fig. \ref{fig:fig13}. Line convention 
as in the left panel of fig. \ref{fig:fig10}.
\label{fig:fig16}
}
\end{figure}

The cross section and the polarization observables calculated for the 
$^{16}$O($\gamma,pp$)$^{14}$C reaction in the out-of-plane kinematics with 
$\phi = 30^{\mathrm{o}}$ are given in figs. \ref{fig:fig13}-\ref{fig:fig16}. 
In comparison with the result for the coplanar kinematics, the cross section in
fig. \ref{fig:fig13} shows the same kinematic effect found in the 
$^{16}$O($\gamma,np$)$^{14}$N reaction. Values of $p_{\mathrm r}$ between 90 and 
160 MeV/$c$ are forbidden, and the region that in the coplanar kinematics of 
fig. \ref{fig:fig10} corresponds to the maximum of $^1S_0$ and to the minimum of
$^3P_1$ is cut. As a consequence, the contribution of $^3P_1$ is always larger 
than the one of $^1S_0$ all over the angular distribution. The size of the final 
cross section is reduced in the region of the maximum and a different shape is 
obtained. The contributions of the one-body and of the two-body current 
displayed in the left panel are of about the same size and add up in the final 
cross section. All the polarization observables in 
figs. \ref{fig:fig14}-\ref{fig:fig16} are sizable. Also in this case both the 
one-body and the two-body current are important. In some observables either 
contribution prevails or appears even dominant in the calculation, but both 
reaction processes are in general important in the final result and their 
theoretical treatment can be tested in combined measurements of the different 
observables. 

Also the polarization observables which vanish in PW are sizable. Their 
determination might be of particular interest for the study of FSI, and in
particular of the correlations between the two outgoing particles in the final 
state, that are neglected in the present model. For the kinematics that are 
usually considered these effects should not be large in the cross section, but
might be emphasized in particular polarization observables.  

\section{Summary and conclusions}
\label{sec:5}
Two-nucleon knockout reactions represent the best suited tool for exploring 
the behaviour of nucleon pairs in nuclei and their mutual interaction. In 
particular, electromagnetic reactions give a direct access to nuclear structure 
and correlations. A combined analysis of cross sections and polarization 
observables is required for this study. 

In this paper we have given the general properties of the 
($\vec{\gamma},\vec{N}N$) reaction where the incident 
photon is linearly or circularly polarized and the polarization of one outgoing 
nucleon is detected. In the most general situation 16 measurable quantities, 
including the cross section, are obtained, each one corresponding to one 
different structure function. These observables allow a complete investigation 
of all the components and the properties of the hadron tensor.

For a complete determination an out-of-plane kinematics is needed, \textit{i.e.} 
a kinematics where one nucleon does not lie on the plane of the photon and of 
the other nucleon. In the more usual coplanar kinematics a situation similar to 
the one of the ($\vec{\gamma},\vec{N}$) reaction is 
recovered, and only 8 observables are nonvanishing. 

Ten  structure functions, and therefore the corresponding observables, vanish
when the PW approximation is considered for the outgoing nucleons. These
quantities appear very well suited to investigate the effect of FSI between the 
ejected nucleons and the residual nucleus, and, possibly, the mutual interaction 
between the two outgoing nucleons. 

Numerical examples have been presented for the exclusive 
$^{16}$O($\gamma,np$)$^{14}$N and $^{16}$O($\gamma,pp$)$^{14}$C 
reactions. $^{16}$O is a well suited target for this investigation. From the 
theoretical point of view, a great deal of work has been done and a large 
experience has been acquired on this nucleus. From the experimental point of 
view, precise measurements have been already carried out for these reactions or 
are planned in different laboratories, aiming at determining cross sections and 
polarization observables for individual final states. It is of particular
interest to consider exclusive reactions. Previous investigations have shown 
that different final states exhibit a different sensitivity to the theoretical 
ingredients of the calculations and to the various reaction processes.
Thus, the experimental resolution of specific final states may act as a filter 
to disentangle and separately investigate different reaction processes and, in 
particular, to study nuclear correlations.

The results are very sensitive to the treatment of the nucleon pair wave 
function and to different aspects involving nuclear structure and correlations. 
The comparison of theoretical predictions with experimental data can thus 
impose severe constraints to the theoretical models. The polarization 
measurements, which are here particularly addressed, are expected to be 
sensitive to the small components of the transition amplitudes, which do not 
show up in the cross sections, where they are generally overwhelmed by the
dominant components.

In this paper cross sections and polarization observables of the 
$^{16}$O($\gamma,np$)$^{14}$N and $^{16}$O($\gamma,pp$)$^{14}$C reactions have 
been calculated with a similar approach and considering the transitions to the 
ground state of the residual nucleus. The nuclear current contains a one-body 
part, which contributes to the transition only through the correlations, and a 
two-body part, given by the pion seagull, the pion-in-flight and the $\Delta$ 
isobar contributions for ($\gamma,np$) reactions, and only by the contribution 
of the $\Delta$ without charge exchange for ($\gamma,pp$) reactions. In the $np$ 
pair wave function the effects of short-range and tensor correlations are 
taken into account, while the $pp$ pair includes a consistent treatment of 
short-range and long-range correlations. 

As an example, a photon energy of 120 MeV, which seems suitable for experiments,
has been chosen in the calculations. At this energy value the $\Delta$ plays 
only a minor role. Thus, in the $^{16}$O($\gamma,np$)$^{14}$N reaction the 
seagull current is dominant. The contribution of the one-body current is not 
very important in general, but can be meaningful in particular situations. The 
contribution of tensor correlations is significant, particularly at high values 
of the recoil momentum. Both  the one-body and the $\Delta$ current are 
important in the $^{16}$O($\gamma,pp$)$^{14}$C reaction. The relevance of either
contribution can be emphasized in particular kinematic conditions and in 
different observables.

The polarization observables are always sizable. They are sensitive to all the
different terms of the nuclear current and also to their interference. 
A combined experimental determination of cross sections and polarizations would 
give a complete information on the reaction process and result in a stringent 
test of the theoretical models. This would make it possible to disentangle the 
different contributions to the reactions and shed light on the genuine nature 
of correlations in nuclei.

\newpage

\begin{table}
\bigskip
\caption[Table I]{
Properties of the structure functions and polarization observables.
\label{tab:sf}
}
\bigskip
\begin{tabular}{|cccccc|}
\hline
&  &  & survives & survives &\\
&  &  & in plane & in PW    &\\
\hline
& $h^u_{11}$ & $\sigma_0$ & yes & yes &\\
& $h'^u_{11}$ & $A_{\mathrm c}$  & no & no &\\
& $h^u_{1-1}$ & $\Sigma_{\frac {\pi} {2}}$ & yes & yes &\\
& $\bar{h}^u_{1-1}$ & $\Sigma_{\frac {\pi} {4}}$ & no & yes &\\
& $h^{\mathrm N}_{11}$ & $P^{\mathrm N}$ & yes & no &\\
& $h'^{\mathrm N}_{11}$ & $P^{\mathrm N}_{\mathrm c}$ & no & yes &\\
& $h^{\mathrm N}_{1-1}$ & $\Sigma^{\mathrm N}_{\frac {\pi} {2}}$ & yes & no &\\
& $\bar{h}^{\mathrm N}_{1-1}$ & $\Sigma^{\mathrm N}_{\frac {\pi} {4}}$ 
& no & no &\\
& $h^{\mathrm{L,S}}_{11}$ & $P^{\mathrm{L,S}}$ & no & no &\\
& $h'^{\mathrm{L,S}}_{11}$ & $P^{\mathrm{L,S}}_{\mathrm c}$ & yes & yes &\\
& $h^{\mathrm{L,S}}_{1-1}$ & $\Sigma^{\mathrm{L,S}}_{\frac {\pi} {2}}$ 
& no & no &\\
& $\bar{h}^{\mathrm{L,S}}_{1-1}$ & $\Sigma^{\mathrm{L,S}}_{\frac {\pi} {4}}$ 
& yes & no &\\
& & & & &\\
\hline
\end{tabular}
\end{table}
\vfil
\vfil\eject

\vfil
\vfil\eject

\vfil
\vfil\eject

\vfil
\vfil\eject

\vfil
\vfil\eject

\vfil
\vfil\eject

\vfil
\vfil\eject

\vfil
\vfil\eject

\vfil
\vfil\eject

\vfil
\vfil\eject

\vfil
\vfil\eject

\vfil
\vfil\eject

\vfil
\vfil\eject

\vfil
\vfil\eject

\vfil
\vfil\eject

\vfil
\vfil\eject

\end{document}